\documentclass[prb,twocolumn,superscriptaddress,footinbib]{revtex4-1} 

\usepackage{longtable}
\usepackage{morefloats}
\usepackage[dvips]{graphicx}
\usepackage{color}
\usepackage{epsfig,graphicx,amsfonts,amsbsy}
\usepackage{amsmath,amsfonts,amsthm,amssymb}
\usepackage{appendix}
\usepackage{makeidx}
\usepackage{url}
\usepackage{verbatim}
\usepackage[bookmarksnumbered,pdfpagelabels=true,plainpages=false,colorlinks=true,linkcolor=blue,citecolor=blue,urlcolor=blue]{hyperref}
\usepackage[rightcaption]{sidecap}
\usepackage{array}
\usepackage{xcolor,cancel}
\usepackage{multirow}
\usepackage{tabularx}
\usepackage{braket} 
\usepackage{dsfont}
\usepackage{amsmath}
\usepackage{ulem}
\usepackage{bbold}
\usepackage{etoolbox}

\newcommand{\tmax}{t_{\scalebox{0.6}{max}}}
\newcommand{\Nph}{N_{\scalebox{0.6}{ph}}}
\newcommand{\ttop}{t_{\scalebox{0.6}{top}}}
\newcommand{\nmax}{\bar{n}_{\scalebox{0.6}{max}}}
\newcommand{\etaef}{\eta_{\scalebox{0.6}{eff}}}
\newcommand{\etaad}{\eta_{\scalebox{0.6}{ad}}}
\newcommand{\asym}{\alpha_{\scalebox{0.6}{as}}}
\begin{document}

\title{Photon pumping in a weakly-driven quantum cavity-spin system}

\author{Christina Psaroudaki}
\affiliation{Department of Physics and Institute for Quantum Information and Matter, California Institute of Technology, Pasadena, CA 91125, USA}
\affiliation{Institute for Theoretical Physics, University of Cologne, D-50937 Cologne, Germany}
\author{Gil Refael}
\affiliation{Department of Physics and Institute for Quantum Information and Matter, California Institute of Technology, Pasadena, CA 91125, USA}
\affiliation{Walter Burke Institute for Theoretical Physics, California Institute of Technology, Pasadena, CA 91125, USA}

\date{\today}
\begin{abstract}
We investigate the photon pumping effect in a topological model consisting of a periodically driven  spin-1/2 coupled to a quantum cavity mode out of the adiabatic limit.  In the strong-drive adiabatic limit, a quantized frequency conversion of photons is expected as the temporal analog of the Hall current. We numerically establish a novel photon pumping phenomenon in the experimentally accessible nonadiabatic driving regime for a broad region of the parameter space. The photon frequency conversion efficiency exhibits strong fluctuations and high efficiency that can reach up $80\%$ of the quantized value for commensurate frequency combinations. We link the pumping properties to the delocalization of the corresponding Floquet states which display multifractal behavior as the result of hybridization between localized and delocalized sectors. Finally we demonstrate that the quantum coherence properties of the initial state are preserved during the frequency conversion process in both the strong and ultra-weak-drive limit. 
\end{abstract}

\maketitle
\section{Introduction} \label{sec:Intro}

The development of efficient frequency conversion mechanisms is a process with various technological applications, relevant for quantum communications and quantum computing \cite{Lauk_2020}. Photonic channels in the IR-band appear to be very attractive for the long-distance transmission of a quantum state, providing low-loss transmission \cite{O'Brien2009} based on quantum-compatible storage and repeater protocols \cite{PhysRevLett.81.5932,PhysRevLett.96.093604,Duan2001}. On the other hand, the elementary quantum processors and memories based on atoms \cite{PhysRevLett.98.193601,Chaneliere2005,Olmschenk486}, spins in quantum dots \cite{Gerardot2008}, or superconducting qubits \cite{Devoret1169}, operate in different frequency regions. Thus, the realization of networks connecting disparate quantum systems \cite{Kimble2008,Tanzilli2005} requires the development of quantum interfaces, capable of bridging the frequency gap \cite{PhysRevLett.68.2153,Kumar:90,Mirhosseini2020}. The bidirectional transfer of quantum information relies on mechanisms that shift a quantum state of light from its original frequency band to a desired one, while preserving all other quantum properties \cite{Andrews2014,Rakher2010}. 
\begin{figure}[b]
	\centering
	\includegraphics[width=1\linewidth]{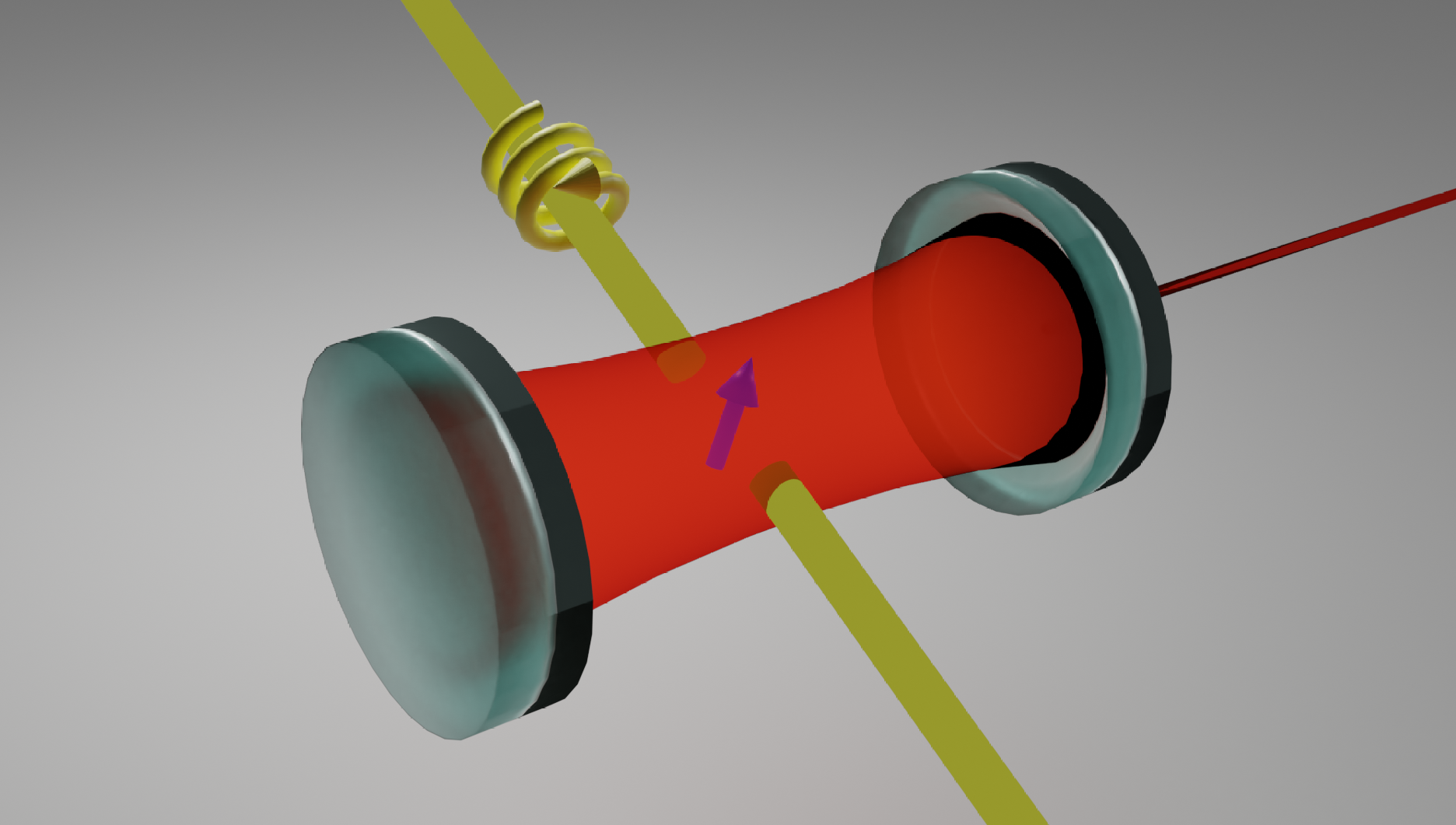}
	\caption{Frequency conversion platform. A single spin (purple) is coupled to a cavity mode of frequency $\omega_1$ (red) and driven by an external periodic magnetic field of frequency $\omega_2$ (yellow). }
	\label{Fig:Spin_White}
\end{figure}

Temporal analogs of topological models can provide a new platform for the efficient transfer of photons from one electromagnetic mode to another. In such systems, novel types of topological order are realized by subjecting a trivial quantum system to periodic driving   \cite{PhysRevB.82.235114,Lindner2011,PhysRevX.6.041001,Choi2017,PhysRevB.97.134303}. In particular, driving a spin-1/2 with two elliptically polarized periodic waves of incommensurate frequencies generates the dynamical analog of a 2D topological insulator, where the Hall current corresponds to a quantized pumping of energy between the two sources \cite{PhysRevX.7.041008,PhysRevB.99.064306}. Recently, this topological frequency conversion effect has been experimentally realized using a single-qubit IBM device \cite{2012.01459}. Analogously, a periodically driven magnetic particle coupled to a cavity induces a quantized energy transfer from the external field to the cavity mode \cite{PhysRevB.99.094311}. Such a quantized frequency conversion occurs as long as the system is in the near-adiabatic limit realized for sufficiently strong drives, when the instantaneous energy gap is large compared to the photon frequencies. Nevertheless, a rough estimation of the single-spin coupling rate to a cavity mode\cite{doi:10.1063/1.3601930} suggests that the strong-drive requirement for a quantized energy flow is experimentally challenging. 

In this work, we investigate the photon pumping effect in a topological model consisted of a periodically driven spin-1/2 coupled to a quantum cavity mode out of the adiabatic limit. The pumping properties of such a non-equilibrium system are described as
the delocalization of the corresponding Floquet states in the synthetic photon-number states along the
direction of the drive frequencies. To describe this phenomenon we employ a statistical analysis of
the Floquet modes using the participation ratio approach, in analogy with electronic transport in the context of Anderson localization\cite{PhysRev.109.1492,Wegner1980}. We thus find echos of Anderson Localization in a dynamics problem. In addition, similarly to Thouless topological charge pumping \cite{PhysRevB.27.6083}, the quantized energy transfer is not expected to remain robust to nonadiabatic effects. Through a numerical estimation of the energy transfer between the two modes, we establish a novel pumping phenomenon that persists in the nonadiabatic driving regime for a broad region of the parameter space of the model. 

We find that the photon pumping efficiency in the weak-driving regime exhibits strong fluctuations. For $\eta \leq \omega_i$, with $\eta$ the driving amplitude and $\omega_i$ the lowest frequency, the strong fluctuations emerge due to the nonadiabatic driving conditions, before the pumping rate saturates to the quantized value for sufficiently strong drives $\eta \approx \etaad$, when adiabaticity is restored. These fluctuations are stronger for rationally-related frequencies, with the conversion efficiency of the order of $80\%$ of the quantized value for $\eta \approx 30\% \omega_1$, while in the same limit irrationally-related frequencies have a vanishing conversion efficiency. In the nonadiabatic limit and around the topological boundary, we demonstrate that a finite pumping phenomenon persists in the trivial phase, but is less effective as we move further away from the boundary. This energy pumping is characterized by a strong sensitivity to the initial phase of the drive,  although we can always specify extended regions of the parameter space with efficiency that exceeds $80\%$ of the quantized value. 

A finite energy pumping signals the delocalization of the corresponding quasienergy states in the space of
photon-number in the cavity. To link the pumping properties of the model to the delocalization of the Floquet states, we follow a statistical analysis of the Floquet modes, closely related to electronic transport in the context of Anderson localization \cite{PhysRev.109.1492,Wegner1980}, applied here to a frequency-space tight-binding model. This is achieved by calculating the inverse participation ratio and extract its scaling exponent with respect to the number of photon states. Our findings corroborate the results of the photon pumping efficiency, namely in the nonadiabatic regime commensurate frequencies have larger delocalization exponents and a finite transport of states in the nontopological regime. Perhaps the most interesting finding is that the Floquet eigenstates also display fractal behavior for any finite $\eta$ as the result of hybridization between the localized and delocalized sectors, induced by the periodic drive. Note that the behavior we explore here is the complement of the non-chaotic frequenecy-locking behavior in the same system investigated in Ref. \onlinecite{PhysRev.Research.2.043411}.

 In addition, we study whether the quantum coherence properties of the initial state are preserved during the frequency conversion process, by evaluating the phase probability of the time-evolved state. During the adiabatic pumping, quantum coherence is preserved and the phase distribution is well-approximated by a Gaussian curve, although the phase undergoes a diffusion process. Surprisingly, in the ultra-weak-drive limit $\eta \ll \omega$, the pumped photon state is described by a central Gaussian peak, suggesting the existence of a state with a well defined phase. As $\eta$ is increased, additional secondary peaks and phase fluctuations are generated and quantum coherence is suppressed.

The structure of the paper is as follows. In Sec.~\ref{sec:QuantEnergy} we discuss the temporal topological models in the strong-drive regime and evaluate the quantized energy pumping for different model parameters. In Sec.~\ref{sec:WeakDrive} we calculate the energy flow in the weak-drive regime and identify regimes of efficient frequency conversion by exploring a broad parameter space. Such an energy flow is  understood in terms of the delocalization of Floquet states, presented in Sec.~\ref{sec:FloquetLoc}, while the coherence properties of the pumped state are investigated in Sec.~\ref{sec:QuantCoher}. The experimental implications of our work are discussed in Sec.~
\ref{sec:Experiment}. Finally, our main conclusions are summarized in Sec.~\ref{sec:Discussion}. 

\section{Quantized Energy Transfer}\label{sec:QuantEnergy} 

The purpose of this section is to present the topological properties of an externally driven spin, coupled to a dynamical quantum mechanical cavity mode and discuss the quantized conversion of energy in the topologically nontrivial parameter regime. The quantum system is described by a Hamiltonian of the form,
\begin{align}
\mathcal{H}(t)= \omega_1 \hat{n} -\eta \hat{\mathbf{B}} \cdot \hat{\boldsymbol{\sigma}} \,,
\label{eq:HamCavity}
\end{align}
where $\hat{\boldsymbol{\sigma}}= (\sigma_x,\sigma_y,\sigma_z)$ and $\hat{\mathbf{B}}= (\hat{B}_x,\hat{B}_y,\hat{B}_z)$, with $\hat{B}_x= B_d \sin(\omega_2 t+\phi)$, $\hat{B}_y=B_0(\hat{a}-\hat{a}^{\dagger})/2i$ and $\hat{B}_z=B_m -B_d \cos(\omega_2 t+\phi)-B_0(\hat{a}+\hat{a}^{\dagger})/2$. Here $\hat{a}$ denotes the photon annihilation operator of the cavity mode with frequency $\omega_1$ and amplitude $B_0$, $\hat{n}\equiv \hat{a}^{\dagger} \hat{a}$ is the photon number operator, $B_m$ represents a static field along the $z$ direction and $B_d$ is the amplitude of the external periodic field of frequency $\omega_2$ and phase $\phi$. We also use $\hbar=1$ throughout. 

 A rough estimation of the single-spin coupling rate to a cavity mode is given by $\eta_0= \eta B_0=g \mu_B \sqrt{\mu_0 \omega_c/2\hbar V_c}$, where $V_c$ is the cavity mode volume and $\omega_c$ the cavity frequency, further enhanced to $\eta_s=\eta_0 \sqrt{N_s}$ by utilizing an ensemble of $N_s$ noninteracting spins \cite{doi:10.1063/1.3601930}. For a microwave cavity mode with $\omega_c=10$ GHz, a small $V_c=10^{-12}$ m$^{3}$ gives $\eta_0=143$ Hz or $\eta=0.32$ GHz for $N_s=5 \times 10^{12}$ spins. The single-spin cavity strength has been experimentally measured using a X-band  resonator to $\eta_0=0.38$ Hz and $\eta=37.4$ MHz for $V_c=2 \times 10^{-10}$ m$^3$ and $N_s=7.8 \times 10^{14}$~ \cite{doi:10.1063/1.3601930}, suggesting that the strong-drive requirement for a quantized energy flow, $\eta \gg \omega_c$, is experimentally challenging. 

The model of Eq.~\eqref{eq:HamCavity} is regarded as a one dimensional semi-infinite tight-binding model, with $n$, the number of photons, being the lattice site. The quantized transfer of energy for incommensurate frequencies occurs in the near-adiabatic limit, when the frequencies $\omega_i$ are smaller than the instantaneous energy. Thus when $\eta \mbox{min}\vert B_c - B_{\pm} \vert \gg \omega_i$, with $B_{\pm}=\vert B_m \pm B_d \vert$ and $B_c = B_0 \sqrt{n}$ the effective amplitude of the electromagnetic mode in a cavity on $n$ photons. Additionally, the system is in its topological regime when $B_c \in[B_-,B_+]$. When the spin is aligned with the instantaneous field, and under the requirements specified above, the topological effect emerges as an increase in the photon number at the quantized rate 
\begin{align}
n_Q = \frac{\omega_2 C}{2 \pi} \,, \label{eq:QuantN}
\end{align}
together with an energy transfer at the rate $dE/dt = \omega_1 \omega_2 C/2 \pi$ with $E=\omega_1 \langle n \rangle$. We note that although the system can be chosen initially to be in its topological regime, since $B_c= B_0 \sqrt{n}$ (with $n$ here a dynamical quantity), we expect that at later times the condition $B_c \in[B_-,B_+]$ breaks down. Thus, an increase of the photon number with a quantized rate can take place as long as $n \in [B_-^2/B_0, B_+^2/B_0]$. 

\begin{figure}[t]
	\centering
	\includegraphics[width=1\linewidth]{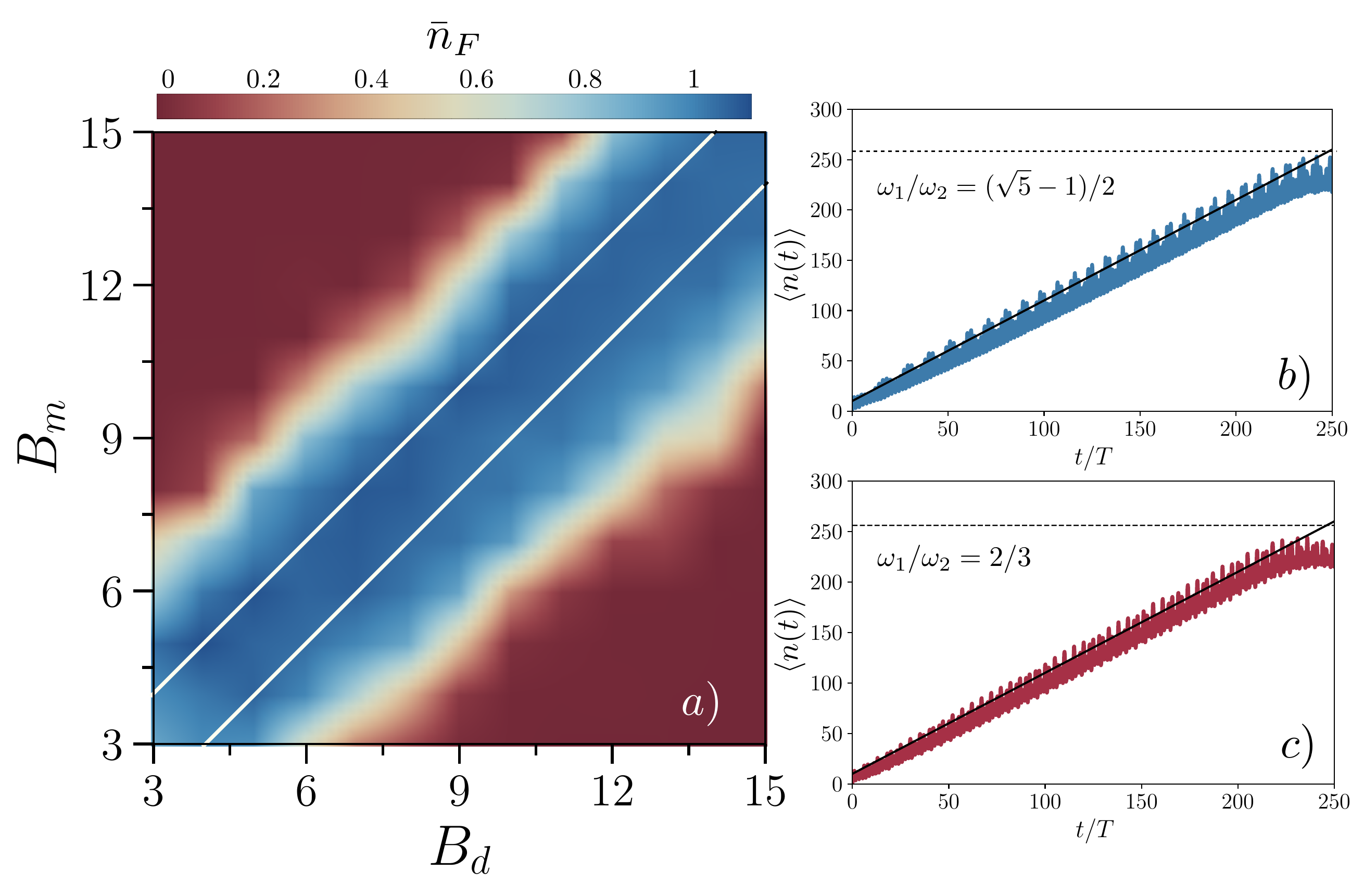}
	\caption{a) Frequency conversion efficiency $\bar{n}_F$ in the strong-drive regime $\eta=2$, between commensurate frequencies $\omega_1/\omega_2=2/3$, as a function of the amplitude of the external drive $B_d$ and the Zeeman field $B_m$. We choose $N_{\mbox{\tiny ph}}=1$, $B_0=1$, $\omega_2=1$, and $\phi=0$. b)-(c) Time evolution of $\langle n(t) \rangle$ for $B_m=8=B_d$ and b) incommensurate frequencies $\omega_1/\omega_2=(\sqrt{5}-1)/2$ with unity efficiency $\bar{n}_F = 1$, and c) commensurate frequencies $\omega_1/\omega_2=2/3$ and efficiency $\bar{n}_F \approx 1$. In both cases, the number of photon in the cavity increases at the quantized rate $n_Q$ depicted with the black solid line from the value $N_{\mbox{\tiny ph}}=1$ until it reaches the topological boundary $(B_m+B_d)^2/B_0=256$ illustrated with black dashed line at approximately time $t/T=(B_m+B_d)^2/B_0-N_{\mbox{\tiny ph}}=255$.}
	\label{Fig:CavitySD}
\end{figure}

\begin{figure}[t]
	\centering
	\includegraphics[width=1\linewidth]{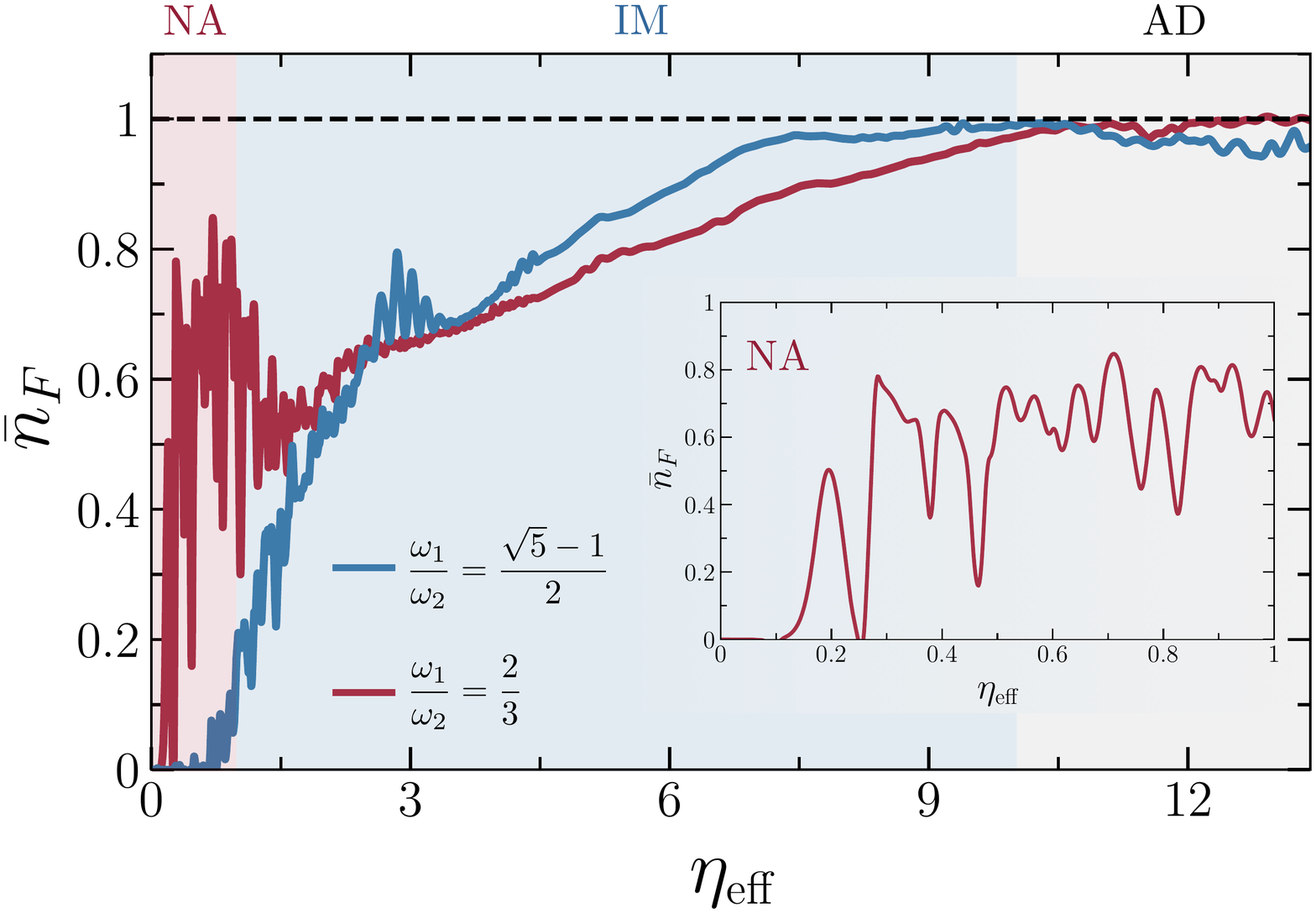}
	\caption{Frequency conversion efficiency $\bar{n}_F$ as a function of the effective drive amplitude $\eta_{\mbox{\tiny eff}}=\eta \sqrt{N_{\mbox{\tiny ph}}}$, for a cavity with initially $N_{\mbox{\tiny ph}}=20$ photons, and a choice of $B_0=1$, $B_m=8=B_d$, $\phi=0$, and $\omega_1/\omega_2=(\sqrt{5}-1)/2$ (blue line) or $\omega_1/\omega_2=2/3$ (red line). Due to the nonadiabatic effects, the efficiency fluctuates before it converges to unity for sufficiently strong drives $\eta_{\mbox{\tiny ad}} \approx 10$. Fluctuations are stronger for commensurate frequencies, with an efficient pumping effect $\bar{n}_F \approx 0.8$ in the deep nonadiabatic regime with $\eta_{\mbox{\tiny eff}}$ as small as $3 \% \times \eta_{\mbox{\tiny ad}}$. }
	\label{Fig:CVComVsIn}
\end{figure}

To address the frequency conversion in the weak-drive or commensurate frequency case, where no quantization should be expected, we numerically calculate\cite{SciPostPhys.2.1.003} the expectation value of the photon number operator as $\langle \hat{n}(t) \rangle= \langle \Psi (t) \vert \hat{n} \vert \Psi(t) \rangle$, with $\vert \Psi(t) \rangle$ the time-evolved state $\vert \Psi(t) \rangle = U(t) \vert \Psi_0 \rangle$ and $U(t) = \mathcal{T} \exp[-i \int_0^{t} \mathcal{H}(t') dt'] $ the time evolution operator \cite{SciPostPhys.2.1.003}. The Hamiltonian of Eq.~\eqref{eq:HamCavity} assumes a quantized light-matter interaction with eigenstates of the form $\vert n,s \rangle$, where $n$ is the photon and $s$ the spin index. At time $t=0$, the system is prepared such that the spin is in its ground state  $\vert\downarrow \rangle$, aligned with the instantaneous magnetic field, and the photon mode in a coherent state with mean number of photons $\Nph$, $\vert \Psi_0\rangle = \vert \Nph, \downarrow \rangle$. With this condition, $\langle \hat{n}(t) \rangle$ is an increasing function over some finite amount of time, while the initial condition $\vert \Psi_0\rangle = \vert \Nph, \uparrow \rangle$ leads to a decreasing photon number (not considered here).

To evaluate the efficiency of the frequency conversion we use the quantized photon increase rate $n_Q$ of Eq.~\eqref{eq:QuantN} as a reference, and study the time-averaged photon number expectation value,
\begin{align}
\bar{n} (\tmax)=\frac{2}{n_Q \tmax^2} \int_0^{\tmax} dt ~(\langle  \hat{n}(t)\rangle-\Nph) \,,
\end{align}
where we used $\langle \hat{n}(t=0)\rangle=\Nph$. Further, we focus on the quantity $\bar{n}_F = \bar{n}(\ttop)$, where $\ttop=T B_+^2/B_0-\Nph$ and $T=2\pi/\omega_2$ the period of the drive. Under this choice, we study the study the frequency conversion effect for $\langle  \hat{n}(t)\rangle \in [B_-^2/B_0, B_+^2/B_0]$. Thus, for unit efficiency $\bar{n}_F=1$, the number of photons in the cavity increases from the value $\Nph$ up to $\Nph + B_+^2/B_0$, at the quantized rate $n_Q$, a process that takes place within a $\ttop$ amount of time. 

The main features of the quantized frequency conversion are depicted in Fig.~\ref{Fig:CavitySD}, plotted in the strong-drive regime $\eta=2$ and a choice of $\Nph=1$, $B_0=1$, $\omega_2=1$ and $\phi=0$. Figs.~\ref{Fig:CavitySD}-b) and c) suggest that in this limit, the number of photons in the cavity increases at the quantized rate $n_Q$, from the initial value $\Nph$ up to the upper topological boundary $B_+^2/B_0$ at time $\ttop$. Choosing $B_m=8=B_d$, we find that incommensurate frequencies $\omega_1/\omega_2=(\sqrt{5}-1)/2$ exhibit unit efficiency $\bar{n}_F = 1$, while commensurate frequencies $\omega_1/\omega_2=2/3$ can be as efficient with $\bar{n}_F \approx 1$. Finally, the colored surface of Fig.~\ref{Fig:CavitySD}-a) represents the conversion efficiency $\bar{n}_F$-dependence on parameters $B_m$ and $B_d$ for the commensurate case $\omega_1/\omega_2=2/3$. White solid lines indicate the boundaries of the topological regime. Simple inspection reveals that there is a broad parameter range beyond the topological regime for which $\bar{n}_F$ reaches and even exceeds the unit efficiency. 

Below we demonstrate that within this parameter region and for the experimentally accessible weak-drive limit, it is possible to transfer photons from the external mode to the cavity mode, starting from the quantum mechanical few-photon limit, with high efficiency that can reach up to $\bar{n}_F=0.8$. 

\section{Nonadiabatic Pumping}\label{sec:WeakDrive}

With this preparation, we turn to the main task of this paper and calculate the frequency conversion efficiency in the weak-drive regime, $\eta \leq \omega_i$. In analogy with the quantization of the charge transport upon a cyclic adiabatic driving of a band insulating system, known as Thouless topological pumping \cite{PhysRevB.27.6083}, the quantized energy transfer is not expected to remain robust in the face of nonadiabatic effects \cite{PhysRevLett.120.106601}. In the adiabatic limit discussed in Sec.~\ref{sec:QuantEnergy}-B, the non-quantized pumping power for commensurate frequencies is understood in terms of a partial sampling of the Berry curvature, contrary to the incommensurate case with a robust quantized pumping proportional to $C=1$. Thus, for an efficient pumping beyond the topological regime, it appears promising to tune the photon frequencies to a rational ratio. At time $t=0$, the system is in the state $\vert \Psi_0 =\vert \Nph,\downarrow \rangle$.    

\begin{figure}[t]
	\centering
	\includegraphics[width=1\linewidth]{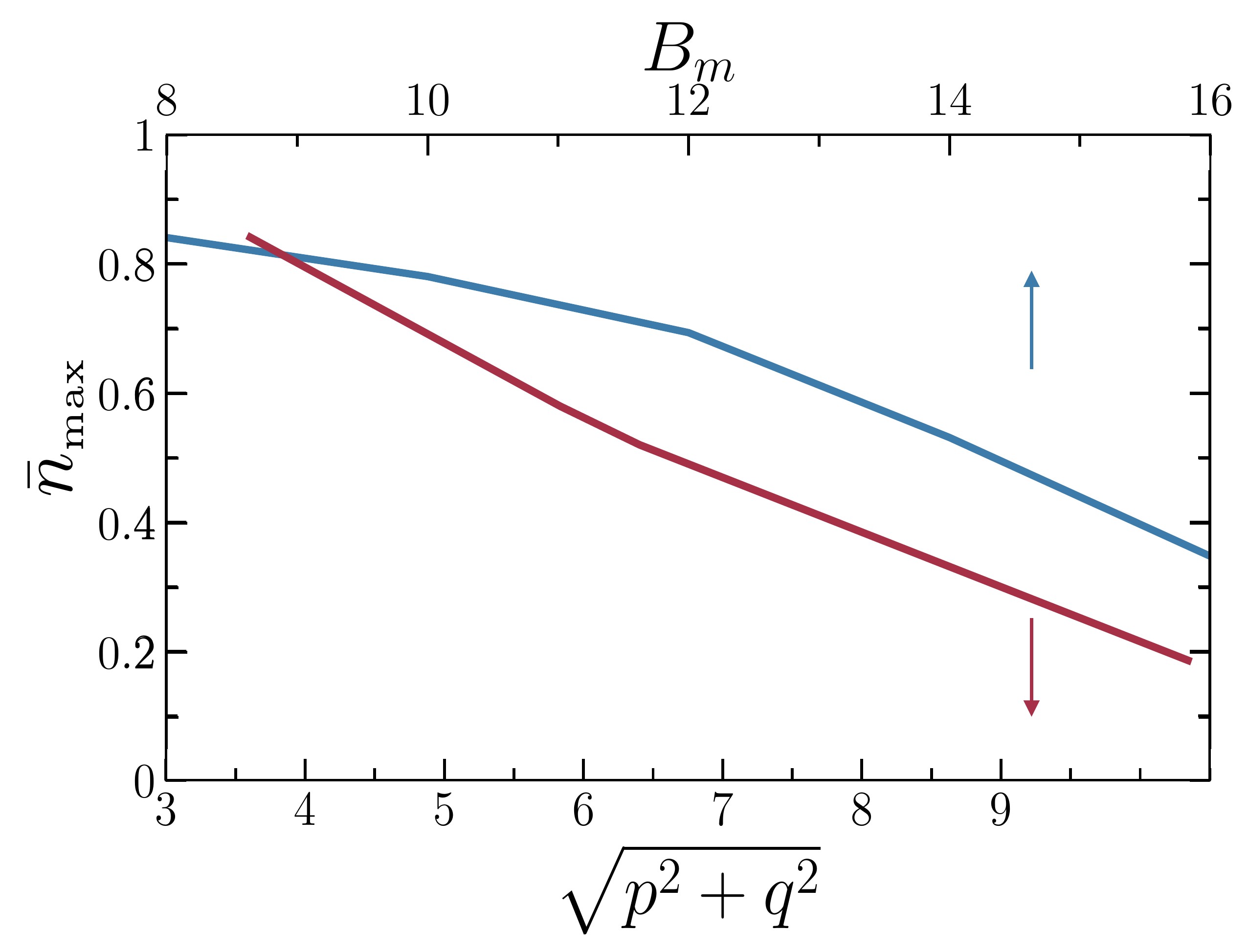}
	\caption{Frequency conversion efficiency $\bar{n}_{\mbox{\tiny max}}$ around the topological transition (blue line), and approaching irrationally-related frequencies (red line). In both cases we use $N_{\mbox{\tiny ph}}=20$, $B_0=1$ and $B_d=8$. The blue line depicts the conversion efficiency as we vary $B_m$ from the topological to the trivial phase for $\omega_1/\omega_2=2/3$ and a topological phase boundary at $B_m=12.5$. The red line depicts the conversion efficiency for $B_m=8$ and rationally-related frequencies of the form $\omega_1/\omega_2 =q/p$, with increasing $\sqrt{p^2+q^2}$.}
	\label{Fig:TopoTrans}
\end{figure}

\begin{figure}[b]
	\centering
	\includegraphics[width=1\linewidth]{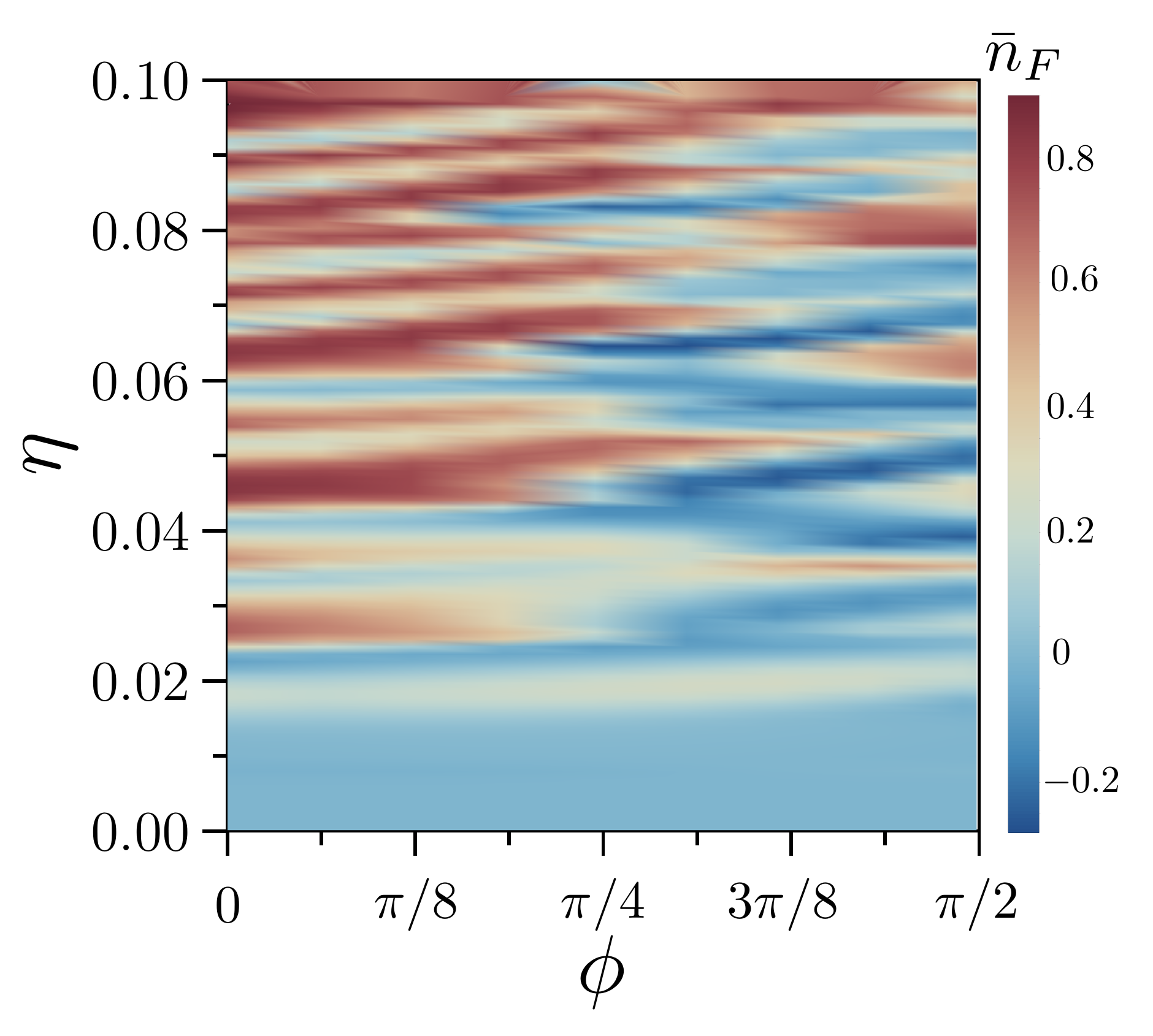}
	\caption{The colored surface represents the conversion efficiency $\bar{n}_F$ as a function of the initial phase of the external periodic drive $\phi$ and the drive amplitude $\eta$, with $B_0=1$, $\omega_1/\omega_2=2/3$, $N_{\mbox{\tiny ph}}=100$ and $B_m=20=B_d$. We observe that $\bar{n}_F$ is highly sensitive with respect to both $\eta$ and $\phi$.}
	\label{Fig:CavityPhase}
\end{figure}

In Fig.~\ref{Fig:CVComVsIn} we plot the frequency conversion efficiency $\bar{n}_F$ as a function of the effective drive amplitude $\etaef=\eta \sqrt{N_{\mbox{\tiny ph}}}$, for a cavity with initially $\Nph=20$ photons in the topological regime with $B_0=1$ and $B_m=8=B_d$. We examine both rational $\omega_1/\omega_2 = 2/3$ (red line) and irrational $\omega_1/\omega_2 = (\sqrt{5}-1)/2$ frequency combinations in a broad region of the driving amplitude including the nonadiabatic (NA) $\etaef \leq \omega_i$ (red shaded area), the intermediate (IM) $\omega_i \lessapprox \etaef \leq \etaad$ (blue shaded area), and the adiabatic regime (AD) (gray shaded are) for $\etaef \geq \etaad$. Fig.~\ref{Fig:CVComVsIn} allows for two major observations. First, for both frequency combinations, due to the nonadiabatic effects, $\bar{n}_F$ strongly fluctuates before it converges to unity for sufficiently strong drives $\etaad \approx 10$.  In addition, in the nonadiabatic regime depicted in detail in the inset of Fig.~\ref{Fig:CVComVsIn}, fluctuations are stronger for rational frequencies, with efficient pumping effects of the order of $\bar{n}_F \approx 0.8$ at $\etaef$ that could be as small as $30 \%  \omega_1$. For intermediate amplitudes $2 \lessapprox \etaef \lessapprox 6$, thus for $\etaef \approx 20-60 \% \times \etaad$, there is still a finite photon pumping $0.5 \lessapprox \bar{n}_F \lessapprox 0.8$, a behavior that is independent of the frequency combination. To justify the choice of the field amplitude $B_m=8=B_d$, we note for the parameters summarized in Fig.~\ref{Fig:CavitySD}-\ref{Fig:CVComVsIn}, the frequency conversion effect takes place within a sufficiently long time interval $\ttop=1071$, during which the cavity photon number is increased by $\Delta n = \langle \hat{n}(\ttop) \rangle -    \langle \hat{n}(0) \rangle=256$ photons. In physical units, $\ttop=1~\mu$s for $\omega_1=1$ GHz, while both $B_m$ and $B_d$ are in the mT regime.

In the nonadiabatic regime $\etaef \leq \omega_i$, irrational frequencies display vanishing efficiency $\bar{n}_F=0$. Such a striking contrast of the nonadiabatic pumping between commensurate and incommensurate frequencies is understood as follows. The energy pumping in the adiabatic regime is generated by engineering synthetic dimensions, where each lattice dimension equals the number of irrationally-related drive frequencies. When the two frequencies have a rational ratio, $\omega_1/\omega_2=q/p$ with $p,q$ coprime integers, the two-dimensional lattice is compactified into a cylinder of circumference $\sqrt{p^2+q^2}$ \cite{PhysRevX.7.041008}. Thus, in the limit of $p,q \gg 1$, one expects the pumping to be insensitive to the details of the frequency ratio and obtain its universal value which derives from the underlying topology. In turn, the topological effect cannot apply at weak drive. This is supported by the results summarized in Fig.~\ref{Fig:TopoTrans} (red line), where we plot $\nmax$ in the nonadiabatic regime for $\Nph=20$, $B_0=1$, $B_m=B_d=8$, and various frequency combinations. Here $\nmax$ is defined as the maximum value of $\bar{n}_F$ in the nonadiabatic regime, $\nmax=\mbox{max}\{ \bar{n}_F(\etaef)	: \etaef\in[0,1]\}$. It becomes apparent that as $p,q$ increase, the nonadiabatic pumping effect vanishes. The full fluctuating behavior of $ \bar{n}_F(\etaef)$ is presented in Fig.~\ref{Fig:CVFreq}. 
\begin{figure}[t]
	\centering
	\includegraphics[width=1\linewidth]{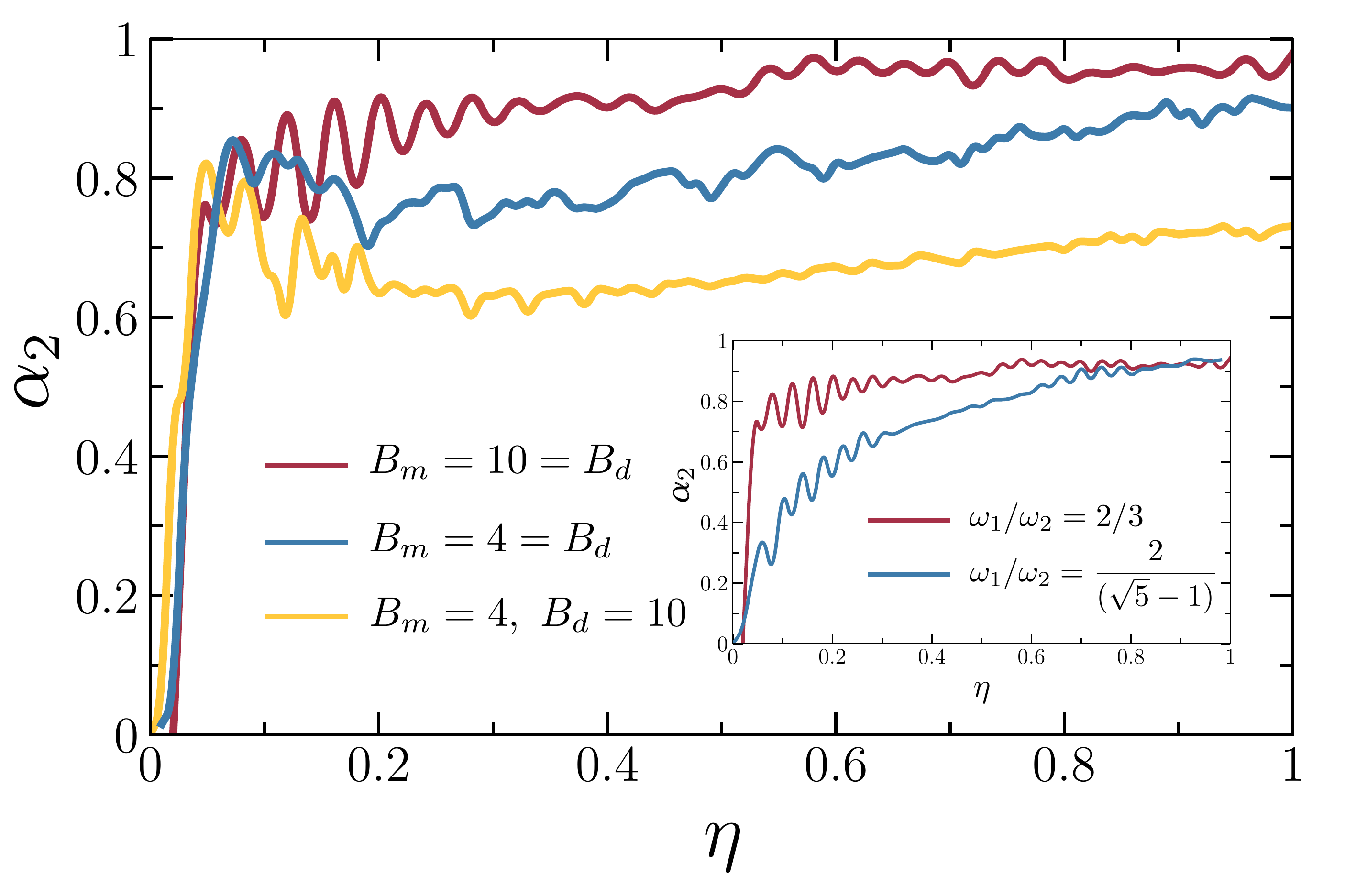}
	\caption{Inverse participation ratio exponent $\alpha_2$ as a function of the driving amplitude $\eta$ for $\omega_1/\omega_2=2/3$ and $B_0=1$. We use three different parameter combinations, two in the topological regime $B_m=10=B_d$ (red line) and $B_m=4=B_d$ (blue line), and one in the trivial regime $B_m=4$, $B_d=10$. The inset depicts $\alpha_2(\eta)$ for $B_0=1$, $B_m=10=B_d$, and $\omega_1/\omega_2=2/3$ (red line) or $\omega_1/\omega_2=2/(\sqrt{5}-1)$ (blue line).}
	\label{Fig:FLocal}
\end{figure}

\begin{figure}[t]
	\centering
	\includegraphics[width=1\linewidth]{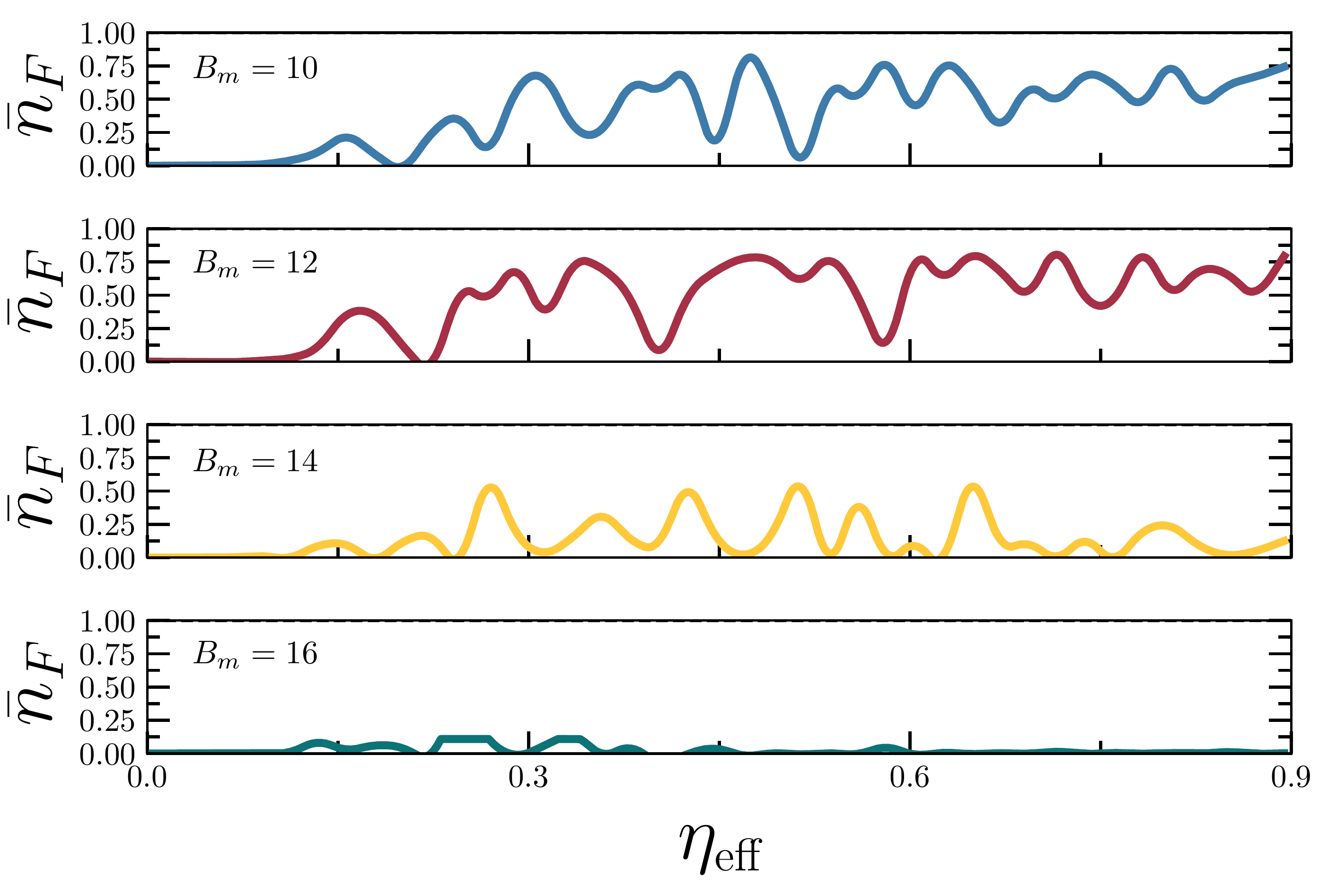}
	\caption{Frequency conversion efficiency $\bar{n}_F$ as a function of the effective drive amplitude $\eta_{\mbox{\tiny eff}}=\eta \sqrt{N_{\mbox{\tiny ph}}}$, for a cavity with initially $N_{\mbox{\tiny ph}}=20$ photons, $B_0=1$, $B_d=8$, and $\omega_1/\omega_2=2/3$. Four different Zeeman fields are presented, two in the topological phase, $B_m=10,12$ and two in the trivial phase $B_m=14,16$. The topological phase boundary is at $B_m=12.47$.}
	\label{Fig:TopoBound}
\end{figure}

Let us now focus on the conversion efficiency around the topological transition, as illustrated in Fig.~\ref{Fig:TopoTrans} (blue line). We present $\nmax$ as a function of $B_m$ for $B_0=1$, $B_d=8$, $\omega_1/\omega_2=2/3$, and a topological phase boundary at $B_m=12.47$. The overall picture suggested is that a finite $\bar{n}_F$ persists in the nontopological regime but with intensity that vanishes as we go further away from the topological boundary. The full fluctuating behavior of $ \bar{n}_F(\etaef)$ is presented in Fig.~\ref{Fig:TopoBound}.

To complete the description we must also examine the sensitivity of $\bar{n}_F$ to the initial phase of the external periodic drive $\phi$, due to dephasing of the time evolved state $\vert \Psi(t) \rangle$. The main features are depicted in Fig.~\ref{Fig:CavityPhase}, where we plot $\bar{n}_F$ as a function of both $\eta$ and $\phi$ and a choice of $B_0=1$, $\omega_1/\omega_2=2/3$, $\Nph=100$ and $B_m=20=B_d$. We observe that although $\bar{n}_F$ is highly sensitive with respect to both $\eta$ and $\phi$, we can specify extended regions of the parameter space with efficient pumping $\bar{n}_F \geq 0.8$. This sensitivity on $\phi$ is a remnant of an analogous $\phi$-dependence of $\bar{n}_F$ in the adiabatic regime, where the pumping effect is proportional to the integral of the Berry curvature along a path selected by $\phi$, further enhanced here by the nonadiabatic driving conditions. The sensitivity of $\bar{n}_F$ on the initial conditions motivates studies on the statistical behavior of the pumping effect for an ensemble of random Hamiltonians. The distribution of the energy pumping efficiency along with the Floquet level statistics is explored in a subsequent publication \onlinecite{Psaroudaki2021}. 

\section{Floquet Localization}\label{sec:FloquetLoc}

A finite energy pumping in a frequency-space tight-binding model signals the delocalization of the corresponding quasienergy states along the direction of the drive frequencies, that now play the role of an effective electric field. In this section we link the pumping properties of the quantum cavity-spin model to the delocalization properties of its Floquet modes. In the case of a simple periodic drive with frequency $\omega_2$, the time evolution of an arbitrary initial state has the form\cite{PhysRev.138.B979}
\begin{align}
\vert \Psi (t) \rangle = \sum_j c_j e^{-i \varepsilon_j t } \vert u_j (t) \rangle \,,
\end{align}
where $\varepsilon_j$ are the quasienergies and $c_j$ are time-independent coefficients. The Floquet states $\vert u_j (t) \rangle$ diagonalize the single-period time evolution operator $U(T)=\mathcal{T}\exp[-i \int_0^{T} \mathcal{H}(t') dt' ]$, where $T=2\pi/\omega_2$ is the period of the drive. 
\begin{figure}[b]
	\centering
	\includegraphics[width=1\linewidth]{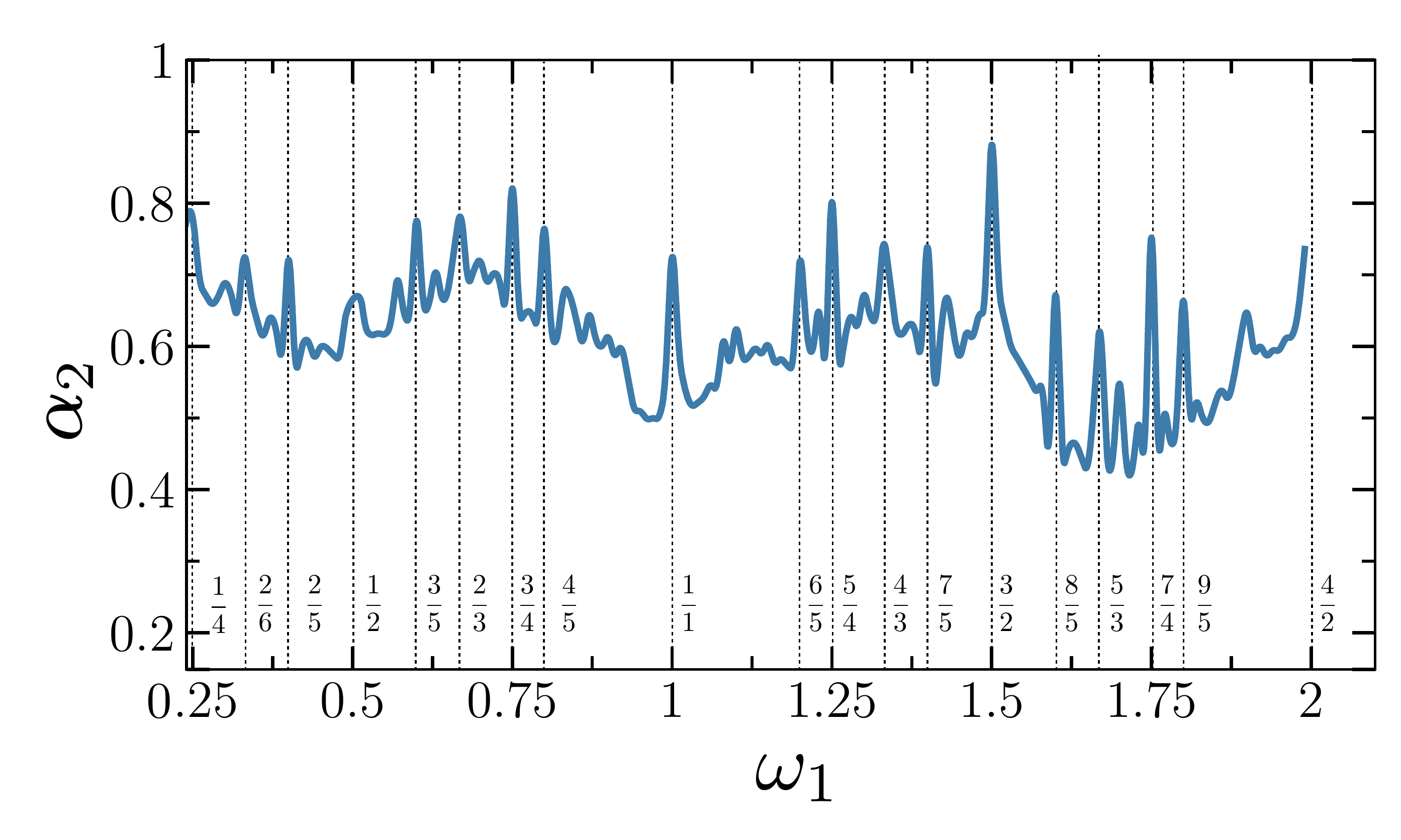}
	\caption{Inverse participation ratio exponent $\alpha_2$ for $B_0=1$, $B_m=10=B_d$, $\eta=0.16$, $\omega_2=1$ and various frequency combinations $\omega_1$. We note that $\alpha_2$ is maximized for commensurate frequencies. }
	\label{Fig:IPNomega}
\end{figure}

Based on an expansion of the zero time Floquet states in the photon basis $\vert u_j (0) \rangle = \sum_n b_j^{n} \vert n \rangle$, the generalized inverse participation ratio (IPR) is defined as \cite{Romito2018}
\begin{align}
\mathcal{R}_q = \frac{1}{N} \sum_{j,n=0}^{N-1} \vert b_j^n \vert ^{2 q} = \frac{1}{N^{\alpha_q}} \,,
\label{eq:IPR}
\end{align}
where $N$ is both the number of the Floquet modes and the number of photons in the cavity, while for convenience we first focus on $\mathcal{R}_2$. If a Floquet state is delocalized among many photon states, then each of them contributes $b_j^n \approx 1/\sqrt{N}$ and $\mathcal{R}_2 \approx 1/N$ for $N \gg 1$ ($\alpha_2=1$). In the opposite limit, a localized Floquet state in the photon lattice will give $\mathcal{R}_2=1$ ($\alpha_2=0$). For a given set of parameters and driving amplitude $\eta$, we calculate $\mathcal{R}_2$ for $N \in [10,200]$ and extract the exponent $\alpha_2$ by a numerical fit.

In Fig.~\ref{Fig:FLocal} we plot the exponent $\alpha_2$ for the model \eqref{eq:HamCavity} as a function of the driving amplitude $\eta$ and examine its behavior as the system transitions from the topological to the trivial regime. In the topological class, Floquet states are delocalized for any finite $\eta$ with $\alpha_2$ exhibiting strong fluctuations in the $0  \leq \eta \leq 0.2$ regime around $\alpha_2 \approx 0.8$, before it asymptotically approaches $\alpha \rightarrow1$ for larger $\eta$. In the trivial class and for large $\eta$, $\alpha_2 \rightarrow \asym$ with $\asym <1$, indicating a hybridization between the localized and delocalized sectors. The inset depicts the delocalization of Floquet states in the topological regime $B_0=1$ and $B_m=10=B_d$ for commensurate $\omega_1/\omega_2=2/3$ (red line) and incommensurate $\omega_1/\omega_2=2/(\sqrt{5}-1)$ (blue line) frequency combinations. As expected, in the nonadiabatic regime commensurate frequencies have larger delocalization exponents, that is translated into large photon pumping in the cavity. This is further supported by the results summarized in Fig.~\ref{Fig:IPNomega} for $B_0=1$, $B_m=10=B_d$, $\eta=0.16$ and various frequency combinations $\omega_1$. We note that $\alpha_2$ is significantly enhanced for commensurate frequency combinations, corroborating the results of Sec.~\ref{sec:WeakDrive}.  

\begin{figure}[t]
	\centering
	\includegraphics[width=1\linewidth]{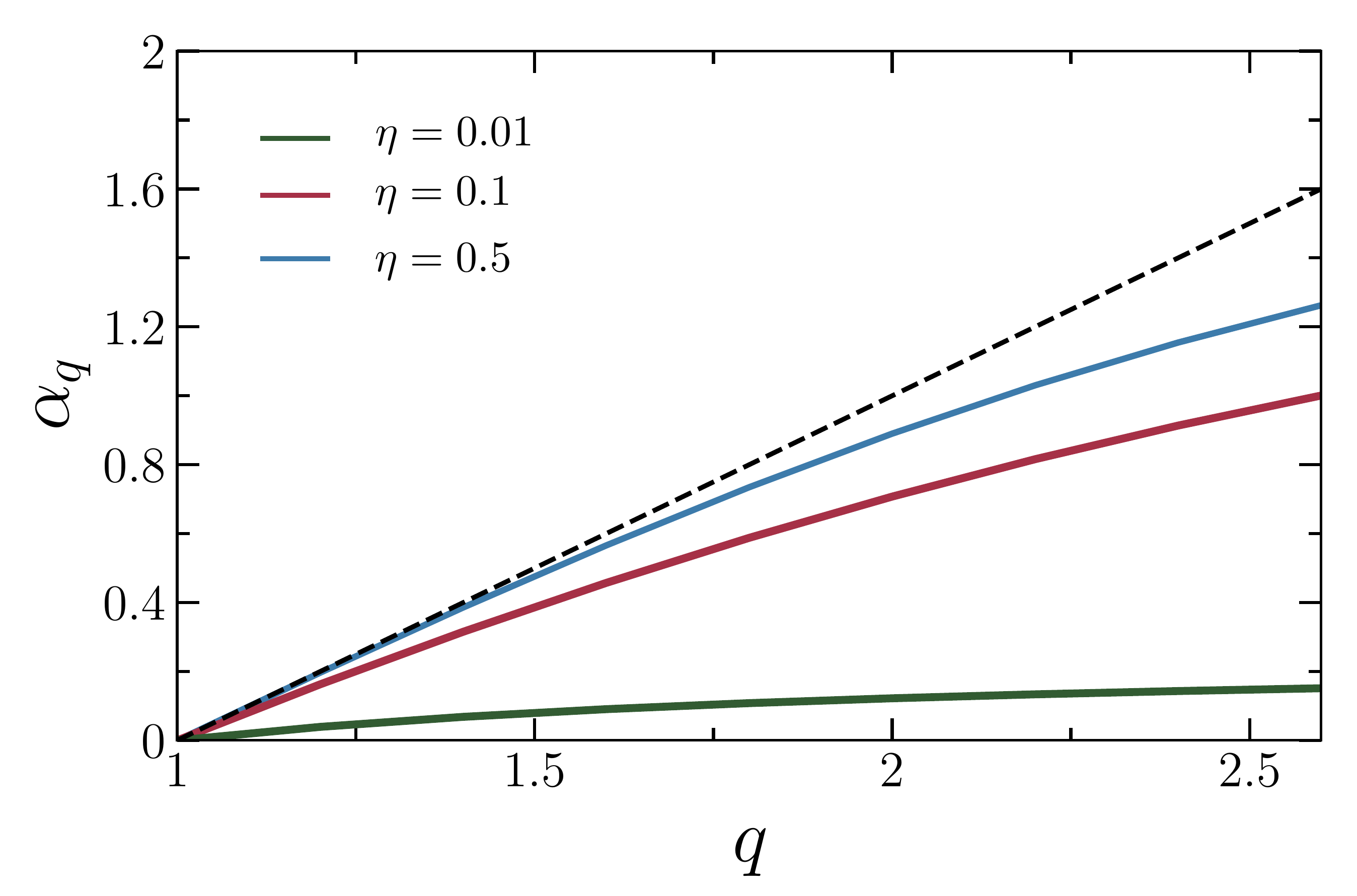}
	\caption{Inverse participation ratio exponent $\alpha_q$ as a function of the fractal dimension $q$ for $B_0=1$, $B_m=10=B_d$, $\omega_1/\omega_2=2/3$ and three driving amplitudes $\eta=0.01,0.1$, and $0.5$. The black dashed line denotes the expected $\alpha=q-1$ behavior for delocalized states. }
	\label{Fig:Fractal}
\end{figure}
Perhaps the most interesting finding is that in all cases the scaling exponent of $\mathcal{R}_2$ reflects a hybridization between the localized and delocalized sectors, that is a sign of Floquet multifractality. The presence of multifractal states is associated with the critical spectral statistics of a system exactly on the metal- insulator transition point \cite{PhysRevB.14.2239} or the Anderson transition point \cite{RevModPhys.80.1355,PhysRevE.54.3221}. To analyze the multifractal behavior of the Floquet eigenstates of the driven quantum-cavity system, we examine the generalized IPR of Eq.~\eqref{eq:IPR} as a function of the fractal dimension $q$. For localized states it holds $\alpha_q=0$, for delocalized $\alpha_q=q-1$, and any other combination signals multifractionality. The overall picture suggested by Fig.~\ref{Fig:Fractal}, where we plot $\alpha_q$ for three values of the drive $\eta=0.01, 0.1,0.5$, is that multifractality of the Floquet states is present for any $\eta>0$ and Floquet states become increasingly delocalized with increasing $\eta$.  This surprising result motivates further studies on the criticality of the Floquet eigenstate statistics \cite{PhysRevLett.102.244102} but is beyond the scope of the present paper and we leave it for the future.

%\subsection{Double-Drive BHZ model}
\section{Quantum Coherence}\label{sec:QuantCoher}

In this section we study the coherent properties of the pumped time-evolved photon state $\vert \Psi(t) \rangle$ to examine whether the phase coherence of the initial state is preserved during the frequency conversion process, a necessary condition for a quantum information transfer \cite{Tanzilli2005}. In the context of quantum optics, a number of probability distributions can be employed to investigate the properties of quantum states. Here we focus on the Pegg-Barnett phase probability distribution based on a construction of a Hermitian phase operator to study whether outgoing states have a well defined phase \cite{PhysRevA.39.1665}. We introduce a complete set of orthonormal phase states \cite{PhysRevA.39.1665}
\begin{align}
\vert \theta_m \rangle = \frac{1}{N} \sum_{n,s} e^{i n \theta_m } \vert n  ,  s \rangle \,, \label{eq:PhaseSt}
\end{align}
where $\theta_m = 2\pi m/N$, $m=0, \cdots ,N-1$, $\vert n, s \rangle $ are eigenstates of the coupled photon-spin system and $N$ is the total number of photon states. The phase probability distribution $P(\theta) = \vert \langle \theta \vert \Psi \rangle\vert ^2$ contains important information on the phase properties of a general state $\vert \Psi \rangle$. Here we initially prepare the system at a coherent photon state $\Nph$ and the spin aligned with the magnetic field, $\vert \Psi_0 \rangle = \vert \Nph , \downarrow \rangle$ and let it evolve under the driven spin-cavity Hamiltonian $\mathcal{H}(t)$ of Eq.~\eqref{eq:HamCavity}. 
\begin{figure}[t]
	\centering
	\includegraphics[width=1\linewidth]{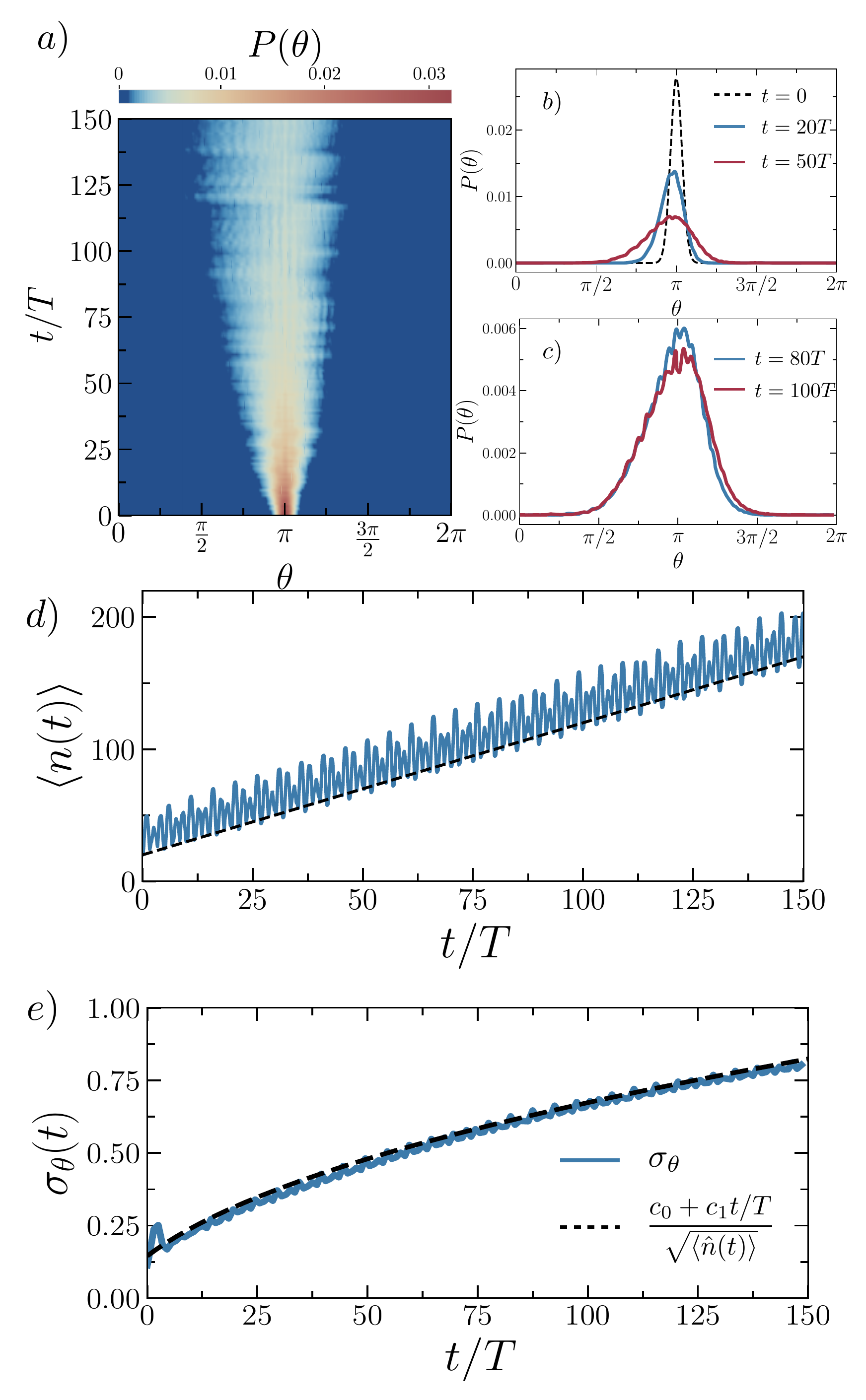}
	\caption{Quantum coherence properties of the pumped photon state in the adiabatic limit $\eta=10/\sqrt{N_{\mbox{\tiny ph}}}$. a) Time evolution of the phase distribution $P(\theta)$ for $B_0=1$, $B_m=8=B_d$ (topological phase), $\omega_1/\omega_2=2/3$, and $N_{\mbox{\tiny ph}}=20$. $P(\theta)$ follows a Gaussian distribution with a standard deviation $\sigma_\theta$ an increasing function over time. b), c) Snapshots of the phase distribution $P(\theta)$ at several time incidents between $t=0$ and $t=100$. d) Time evolution  of $\langle n(t) \rangle$ with $\bar{n}_F =1$. e) Time evolution of $\sigma_\theta$, well approximated by $\sigma_\theta (t) = (c_0 +c_1 t/T)/\sqrt{\langle \hat{n}(t) \rangle}$, with $c_0=0.65$ and $c_1=0.068$.}
\label{Fig:Adiab}
\end{figure}
\begin{figure}[t]
	\centering
	\includegraphics[width=1\linewidth]{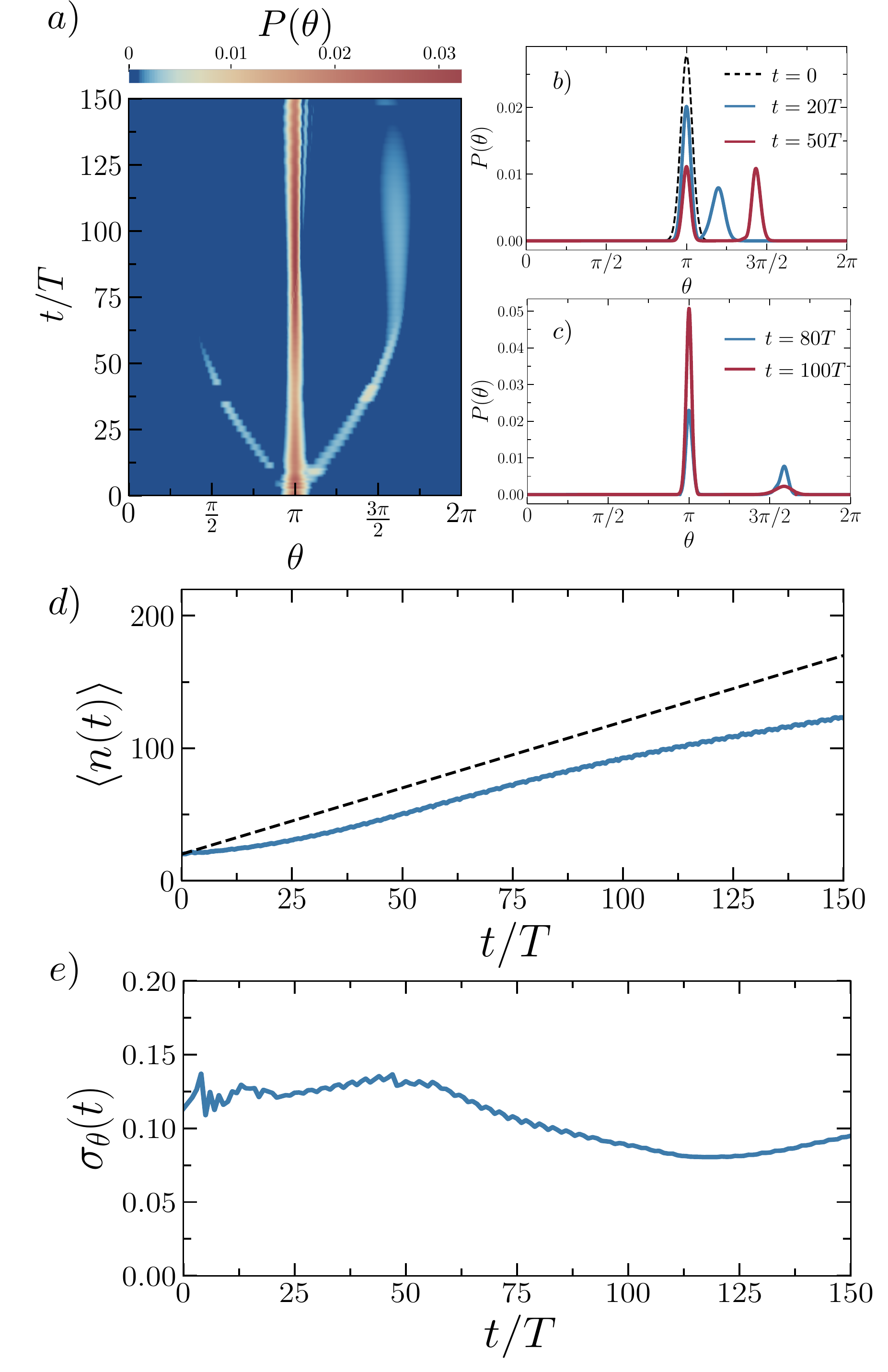}
	\caption{Quantum coherence properties of the pumped photon state in the nonadiabatic limit $\eta=0.3/\sqrt{N_{\mbox{\tiny ph}}}$. a) Time evolution of the phase distribution $P(\theta)$ for $B_0=1$, $B_m=8=B_d$ (topological phase), $\omega_1/\omega_2=2/3$, and $N_{\mbox{\tiny ph}}=20$. b),  c) Snapshots of the phase distribution $P(\theta)$ at several time incidents between $t=0$ and $t=100$. d) Time evolution of $\langle n(t) \rangle$ with $\bar{n}_F \approx 0.8$. e) Time evolution of $\sigma_\theta$ of the central Gaussiann peak.}
	\label{Fig:NAdiab}
\end{figure}

The quantum coherence properties of the pumped photon state $\vert \Psi(t) \rangle$ in the adiabatic limit are summarized in Fig.~\ref{Fig:Adiab}, for parameters in the topological phase $B_0=1$, $B_m=8=B_d$, $\omega_2=1$, commensurate frequencies $\omega_1/\omega_2=2/3$, and strong drive $\eta = 10/\sqrt{\Nph}$. Panel a) depicts the time evolution of the Pegg-Barnett phase distribution $P(\theta)$ for $\Nph=20$. As expected quantum coherence is maintained and $P(\theta)$ is described by a Gaussian peak at all times [see Fig.~\ref{Fig:Adiab}-b) and c)], although the standard deviation $\sigma_\theta$ of the distribution increases over time [see Fig.~\ref{Fig:Adiab}-e)]. The photon number $\langle n(t) \rangle$ increases over time at the quantized rate $n_Q$ [see Fig.~\ref{Fig:Adiab}-d)]. The standard deviation $\sigma_\theta$ is well approximated by $\sigma_\theta(t) = (c_0+c_1 t/T)/\sqrt{\langle n(t) \rangle}$ with $c_0=0.65 $, $c_1= 0.068$, and $T=2\pi/\omega_2$.

\begin{figure}[t]
	\centering
	\includegraphics[width=1\linewidth]{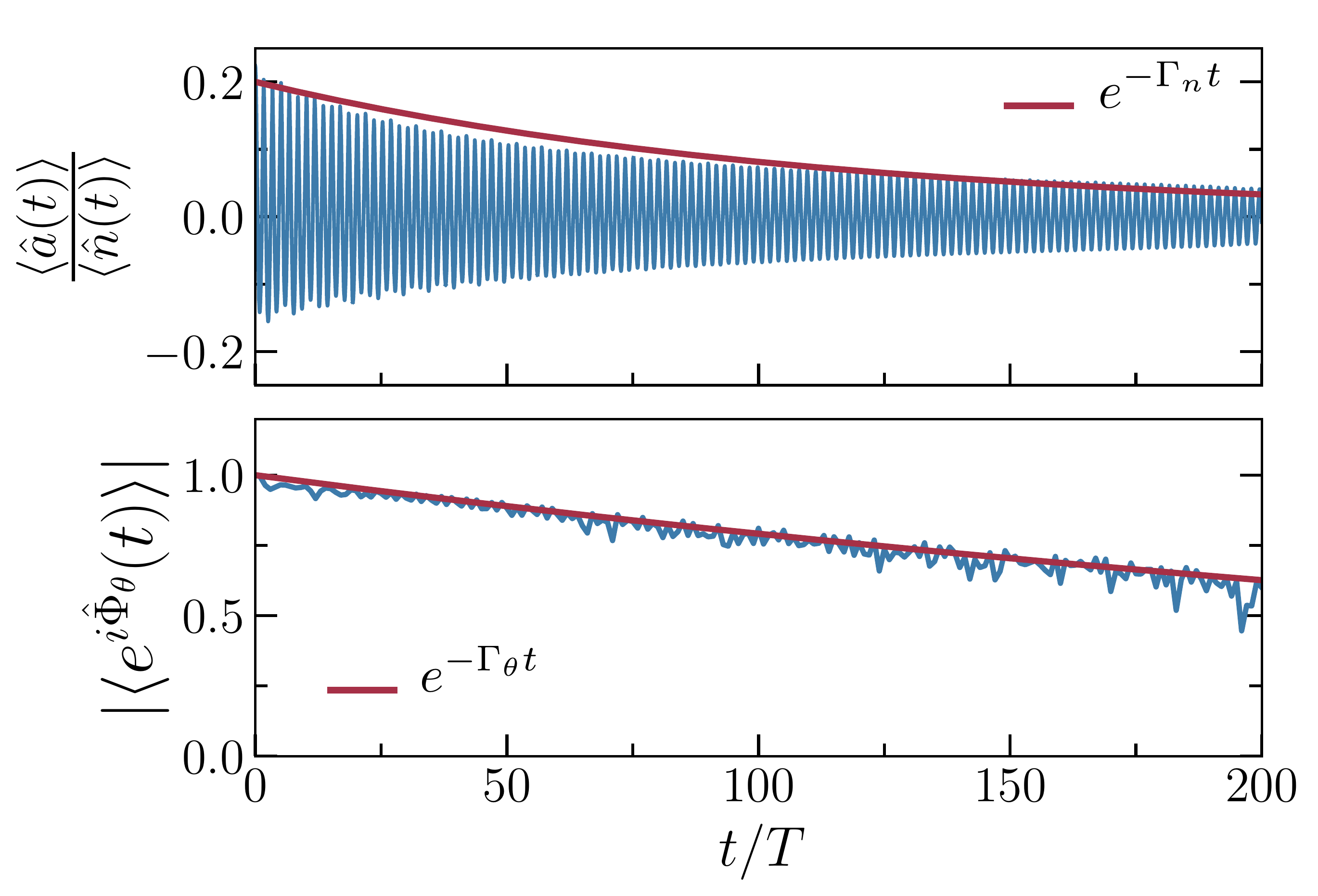}
	\caption{Time evolution of the mean amplitude $\langle \hat{a}(t)\rangle/\langle \hat{n}(t)\rangle$ (upper panel) and $\langle e^{i \hat{\Phi}_\theta} (t)\rangle$ (lower panel) for $\eta=10/\sqrt{N_{\mbox{\tiny ph}}}$, $B_0=1$, $B_m=8=B_d$, $\omega_1/\omega_2=2/3$, and $N_{\mbox{\tiny ph}}=20$. The rate of the amplitude decay is $\Gamma_n = 0.011$ for the former and $\Gamma_\theta= 0.0023$ for the later. We note that on top of the exponential behavior,  $\langle \hat{a}(t)\rangle/\langle \hat{n}(t)\rangle$ oscillates at a frequency $\omega_1 =2/3$. }
	\label{Fig:Difussion}
\end{figure}

The broadening of the Gaussian curve suggests that the phase of the pumped state undergoes a diffusion process, evident in the time evolution of both the mean amplitude $\langle \hat{a}(t) \rangle$ and $\langle e^{i \hat{\Phi}_\theta} (t)\rangle$. Here $\hat{a}$ is the photon annihilation operator and $\hat{\Phi}_\theta = \sum_m \theta_m \vert \theta_m \rangle \langle \theta_m \vert$ is the phase operator. In Fig.~\ref{Fig:Difussion} we present $\langle \hat{a}(t) \rangle/\langle \hat{n}(t) \rangle$ (upper panel) and extract the rate of amplitude decay as $\Gamma_n = 1.7\times 10^{-3}$, a direct measure of the phase diffusion rate \cite{walls_milburn_1995}. The lower panel depicts $\vert \langle e^{i \hat{\Phi}_\theta} (t) \rangle \vert = e^{-\frac{1}{2} \sigma^2_\theta(t)} \approx e^{-\Gamma_\theta t}$ with $\Gamma_\theta = c_1^2/2T=3.66\times 10^{-4}$. Thus, the outgoing pumped state is described by a diffusion coefficient $\Gamma_\theta \approx 3.66~ \times 10^{-4} \times \omega_1$, a quantity that is equivalent to a single-mode laser linewidth. Typically, narrow-band sources are preferable for quantum information applications as they provide access to atomic transitions.   

The quantum coherence properties of the pumped photon state $\vert \Psi(t) \rangle$ in the nonadiabatic limit are summarized in Fig.~\ref{Fig:NAdiab}, for parameters in the topological phase $B_0=1$, $B_m=8=B_d$, commensurate frequencies $\omega_1/\omega_2=2/3$, and ultra-weak drive $\eta = 0.3/\sqrt{\Nph}$. Panel a) depicts the time evolution of the Pegg-Barnett phase distribution $P(\theta)$ for $\Nph=20$. $P(\theta)$ is still described by a central Gaussian peak [see Fig.~\ref{Fig:Adiab} b) and c)], although additional peaks develop due to the nonadiabatic driving conditions. Thus, the standard deviation $\sigma_\theta$, plotted in Fig.~\ref{Fig:NAdiab}-e), increases/decreases over time due to the appearance/disappearance of secondary peaks. For all cases plotted, the photon number in the cavity increases over time with frequency conversion efficiency $\bar{n}_F\approx 0.8$. As we further increase $\eta$ additional phase fluctuations are generated and phase coherence is suppressed. A summary of this behavior is depicted in Fig.~\ref{Fig:NAToA} of the Appendix, where we also illustrate how phase coherence is restored as the system approaches the strong-drive adiabatic limit.  

In summary, the frequency conversion in the adiabatic limit preserves the quantum coherence properties of the initial state, although the state is characterized by a diffusion coefficient. For nonadiabatic driving conditions and ultra-low drive, $P(\theta)$ is still described by a central Gaussian peak, suggesting that the pumped quantum state has a well defined phase. As we further increase the driving amplitude, additional secondary peaks and phase fluctuations are generated and quantum coherence is suppressed. 

\section{Experimental Implications}\label{sec:Experiment}

Here we discuss whether the nonadiabatic frequency conversion discussed in the previous section is experimentally feasible. To give an estimate in physical units under the choice of parameters summarized in Fig.~\ref{Fig:CVComVsIn}, thus for $B_0=1$ and $B_d=8=B_m$, we first note that the adiabatic requirement holds as long as $\eta \geq \hbar \omega_1\etaad /\sqrt{\Nph} \approx 10 \hbar \omega_1 /\sqrt{\Nph}$. Therefore, in the few-photon limit, starting with a state of $\Nph=20$ \textit{coherent} photons, the adiabatic requirement is met for a spin-photon coupling of the order of $\eta/h\omega_1= 0.18$ at $B_d=B_m=100$ mT, and grows to $\eta/ h \omega_1 = 0.8$ in the fully quantum limit with $\Nph=1$ and $B_d=B_m=450$ mT. In the opposite many-photons limit, starting with a cavity of $\Nph=200$ photons, the adiabatic limit corresponds to $\eta/h\omega_1= 0.06$ at $32$ mT, and for $\Nph=400$ we find $\eta/h\omega_1= 0.04$ at $23$ mT. On the contrary, a nonadiabatic pumping is possible for even smaller values of the spin-photon coupling of approximately $\eta=0.3 \hbar \omega_1 /\sqrt{\Nph}$. This translates to $\eta/h \omega_1 = 5.3 \times 10^{-3}$ for $\Nph =20$ at $3$ mT, and $\eta/h \omega_1 = 2.4 \times 10^{-2}$ for $\Nph =1$ at $14$ mT. 

The direct interaction of a single spin with the cavity magnetic field is exceedingly small, with $\eta_0/h\omega_1 =10^{-9}$. A  generalization is to consider of an ensemble of $N_s$ identical two-level systems resonantly interacting with a single electromagnetic field, a model analyzed by Tavis and Cummings \cite{PhysRev.170.379}. In the scope of this model, the collective spin-photon coupling is $\eta=\sqrt{N_s} \eta_0$, where $\eta_0$ is the coupling strength of each individual spin, a scaling that has been experimentally verified, while a similar $\sqrt{N_s}$ behavior prevails also for the magnon-photon coupling in magnetically interacting systems\cite{PhysRevLett.111.127003,PhysRevLett.113.083603,PhysRevLett.113.156401,PhysRevB.94.054433,doi:10.1063/1.4941730}. The coupling of large electron spin ensembles to microwave cavity photons has been experimentally found to be of the order of $\eta/h\omega_1=5\times 10^{-4}$ ($\omega_1 =9.7$ GHz) for a large ensemble of $N_s=10^{16}$ spins in a 3D microwave cavity\cite{doi:10.1063/1.3601930}, and of the order of $\eta/h\omega_1=4 \times 10^{-3}$ for $N_s=10^{12}$ spins in a superconducting cavity\cite{PhysRevLett.105.140502,PhysRevLett.105.140501} ($\omega_1=2.87$ GHz at $40$ mK). 

Several approaches have been explored in order to enhance the spin-photon coupling. Recently, it has been demonstrated that by reducing the cavity effective volume at superconducting nanoconstrictions, the microwave magnetic field is enhanced and results strong spin-photon couplings of the order of $\eta/h\omega=1.4 \times 10^{-3}$ for $N_s = 10^{8}$ spins in a microwave cavity with an average of $5\times 10^5$ thermal photons\cite{Gimeno2020} ($\omega_1=1.4$ GHz at $44$ mK). A different approach is to consider a number of exchange-coupled spins, where the coupling strength is reported to be at least one order of magnitude larger as compared to noninteracting spins, \textit{i.e.} $\eta/h\omega_1=7.6\times 10^{-2}$~ \cite{PhysRevLett.111.127003} ($\omega_1=5.6$ GHz at $50$ mK). In these systems, microwave cavity photons interact coherently with the collective spin excitations in ferromagnetic crystals \cite{doi:10.1063/1.4941730}, such as the ferromagnetic insulator yttrium iron garnet (YIG). A strong coupling is achieved when the cavity and the magnetostatic mode are on resonance, leading to $\eta/h \omega_1 \approx 3 \times 10^{-3}$ ~\cite{PhysRevLett.113.083603} ($\omega_1=10.6$ GHz at $10$ mK). The resonance condition justifies the single photon mode assumption in the model of Eq.~\ref{eq:HamCavity}, as higher-energy modes are off-resonant and can be neglected. This is further supported by the results of Ref.~\onlinecite{PhysRevB.99.094311} in a classical multi-mode cavity, where the authors demonstrated that the energies of higher-energy modes, effectively suppressed by tuning their dissipation rates, decay fast to zero. In view of the increasing interest on the coherent interaction between magnons and microwave or optical photons, we believe our work could serve as a basis for studies on photon frequency conversion in a system of $N_s$ \textit{interacting} spins, also motivated by recent results on the enhancement of frequency conversion in a system of two interacting spins \cite{PhysRevResearch.2.022023}.  

Nevertheless, in the above studies the spin-phonon coupling has been explored away from the quantum single-spin to single-photon limit. A promising platform for the realization of the Tavis-Cummings model is the use of two-state atoms coupled to a resonant cavity mode with a large reported single-qubit to single-photon coupling of the order of $\eta_0/h \omega_1=1.3 \times 10^{-2}$ ~\cite{PhysRevLett.103.083601}$\omega_1=6.7$ GHz at $20$ mK). The coupling of single electron spins in silicon quantum dots to single microwave photons has been reported to be of the order of $\eta_0/h \omega_1 = 2-7 \times 10^{-3}$, with an average thermal photon number in the resonator well below 1\cite{Mi156,Mi2018,Landig2018} ($\omega_1=7.7$ GHz at $10-30$ mK). Finally, at room temperature, the magnon to microwave photon coupling is of the order of $\eta/h \omega_1=1.4-3.3 \times 10^{-3}$ ~\cite{PhysRevLett.113.156401,PhysRevB.94.054433} ($\omega_1=7.9$ GHz), while the spin-photon coupling in a molecular crystal is $\eta/h\omega_1=1.2\times 10^{-3}$ ~\cite{Breeze2017} ($\omega_1=1.45$ GHz). From the above considerations it becomes apparent that the predicted frequency conversion effect in the nonadiabatic limit is experimentally feasible using a large ensemble of spins and is within experimental reach for the quantum few-photon limit. 
 
\section{Discussion}\label{sec:Discussion}

In this work, we consider the transfer of energy in a periodically driven spin-1/2 coupled to a quantum cavity mode out of the adiabatic limit. We establish a novel pumping phenomenon that persists in the weak-drive regime and examine its efficiency for a broad range of the parameter space. We demonstrate that the frequency conversion efficiency exhibits strong fluctuations due to the nonadiabatic effects before it saturates to unity for sufficiently strong drives, $\etaad \approx 10 \omega_i$. It is more efficient for rationally-related frequencies and can reach up to $80\%$ of the quantized value for $\eta = 0.3 \omega_i$. Emphasis is put on the magnitude of the pumping effect in the nontopological phase of the model. A finite frequency conversion efficiency persists in the trivial phase, but is less effective as we move further away from the topological boundary.

A finite-energy pumping signals the delocalization of the corresponding Floquet states, which are found to display a multifractal behavior, due to the hybridization between localized and delocalized sectors. The presence of multifractal states, associated with critical spectral statistics, motivates future studies on the criticality of the Floquet eigenstate statistics. Finally, we show that in the adiabatic limit, during the frequency conversion process, the quantum coherent properties of the initial state are preserved, but the phase undergoes a diffusion process. For ultra-low drive, the pumped quantum state has still a well defined phase but as we further increase the driving amplitude, additional secondary peaks and phase fluctuations are generated and quantum coherence is suppressed. 

Experimental systems are subject to dissipation mechanisms that need to be incorporated for the construction of a realistic model (although we don't expect a qualitative change of our results). In the adiabatic driving limit and within a Markov-Lindblad framework, the inclusion of external noise and dissipation stabilizes the conversion effect as the system approaches a steady state with a quantized number of emitted photons per driving period \cite{PhysRevB.99.094311}. In the nonadiabatic limit, it was recently demonstrated that the introduction of a non-Hermitian  tailored time-periodic dissipation restores the topological transport quantization of Thouless pumps in plasmonic waveguide arrays, emphasizing the uniqueness of Floquet topological systems \cite{Fedorova2020}. 
\begin{acknowledgments} We are grateful to Ivar Martin and Frederik Nathan for useful discussions. C.P. has received funding from the European Union's Horizon 2020 research and innovation program under the Marie Sklodowska-Curie grant agreement No 839004. We are also grateful to the U.S. Department of Energy, Office of Science, Basic Energy Sciences under  Award  de-sc0019166. GR  is  also  grateful  to  the NSF DMR grant number 1839271.  NSF and DOE supported  GR’s  time  commitment  to  the  project  in  equal shares. NSF provided partial support to C.P. This work was performed in part at Aspen Center for Physics, which is supported by National Science Foundation grant PHY-1607611. G.R. is also grateful for support from the Simons Foundation and the Packard Foundation. 

\end{acknowledgments}

%This novel pumping phenomenon offers new possibilities for an efficient frequency conversion using temporal topological models in the experimentally accessible nonadiabatic regime. It also motivates further studies using nontopological models, for example spin waves and magnetostatic modes, that are particularly attractive for the microwave up-conversion of photons due to the large magneto-photon coupling \cite{PhysRevLett.104.077202}.
\begin{figure}[t]
	\centering
	\includegraphics[width=1\linewidth]{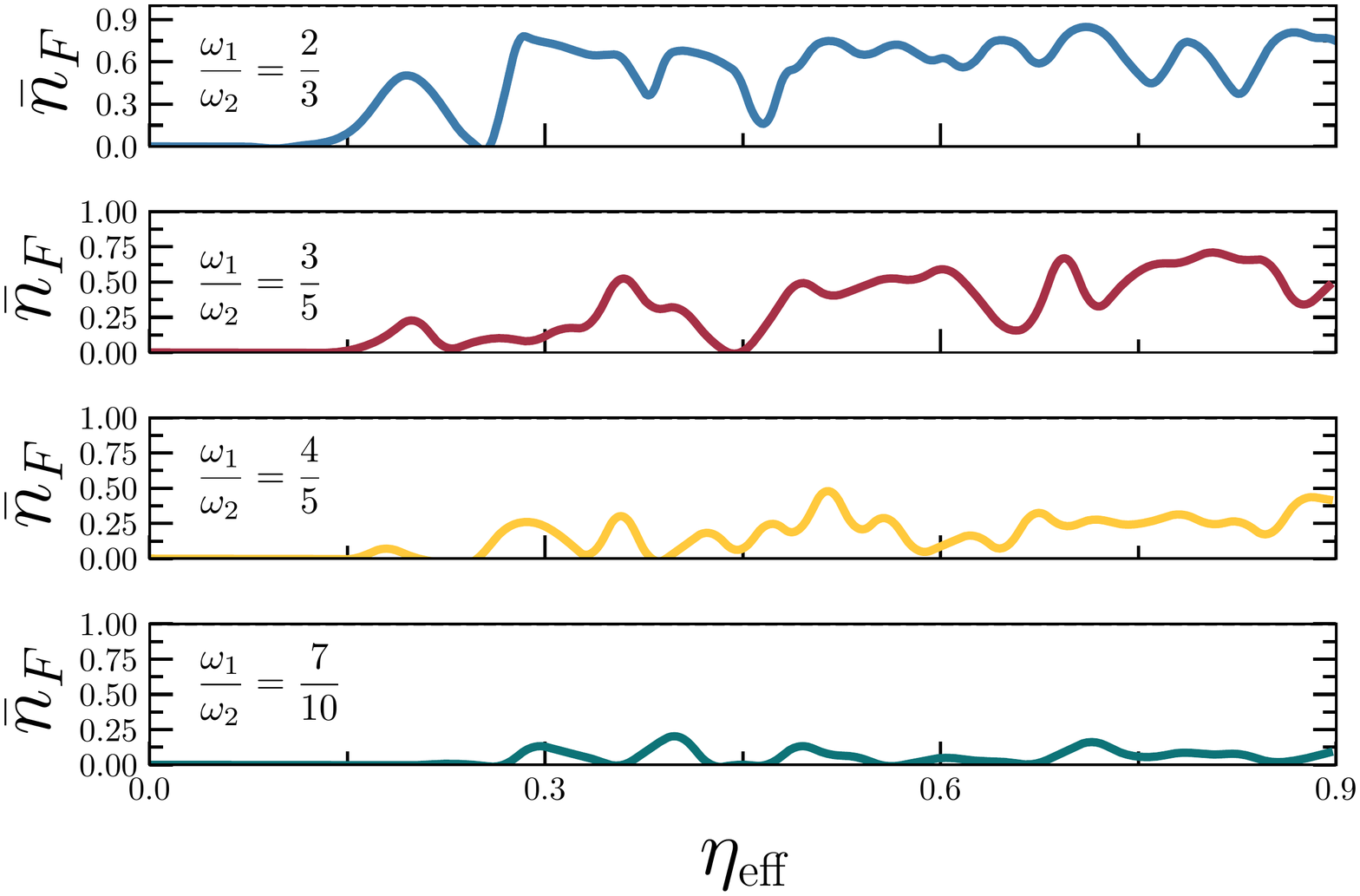}
	\caption{Frequency conversion efficiency $\bar{n}_F$ as a function of the effective drive amplitude $\eta_{\mbox{\tiny eff}}=\eta \sqrt{N_{\mbox{\tiny ph}}}$, for a cavity with initially $N_{\mbox{\tiny ph}}=20$ photons in the topological class $B_0=1$, and $B_m=8=B_d$. Four different commensurate frequency combinations are presented of the form $\omega_1/\omega_2 =q/p$, with increasing $\sqrt{p^2+q^2}$. When $q,p \gg 1$, the nonadiabatic photon pumping vanishes.}
	\label{Fig:CVFreq}
\end{figure}

\begin{figure*}[t]
	\centering
	\includegraphics[width=1\linewidth]{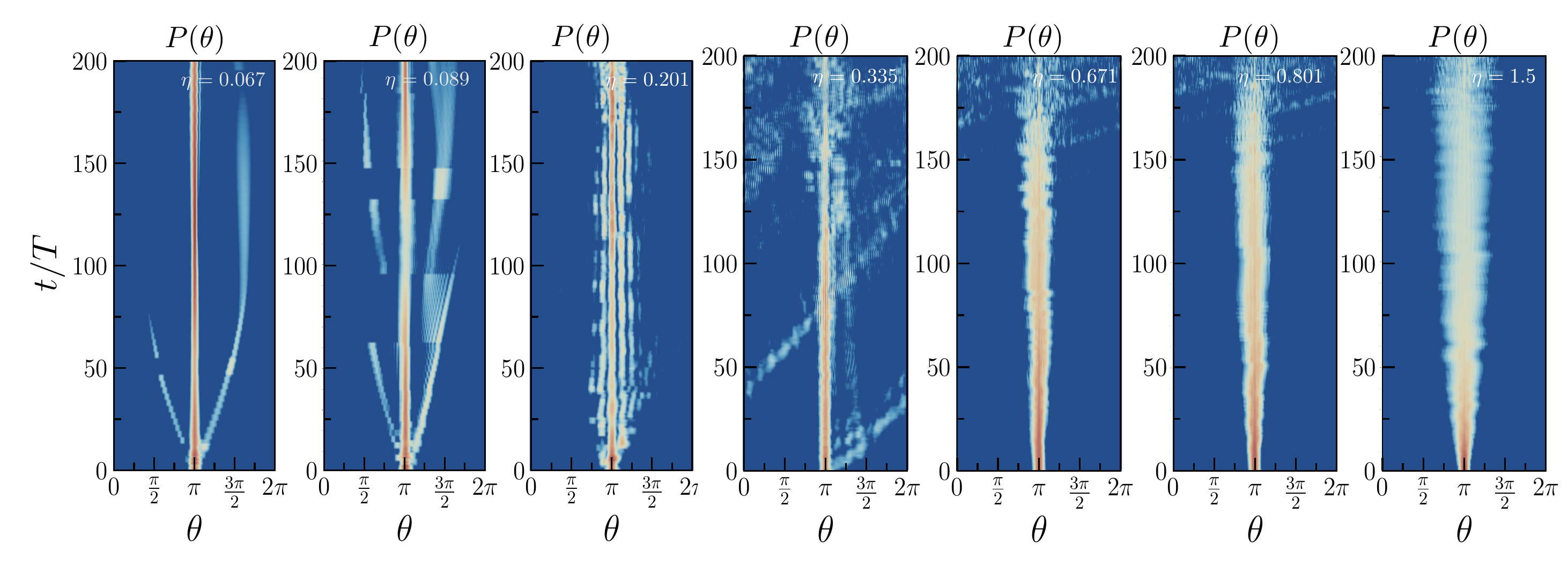}
	\caption{Time evolution of the phase distribution $P(\theta)$ for parameters ranging between the ultra-weak-drive limit $\eta=0.067$ up to the strong-drive limit $\eta = 2.236$. We use $B_0=1$, $B_m=8$, $B_d=8$, $N_{\mbox{\tiny ph}}=20$, and $\omega_1/\omega_2=2/3$. In the ultra-weak limit $\eta \ll \omega_i$ (left panel), $P(\theta)$ is described by a central Gaussian peak, suggesting that the phase is preserved during the photon pumping effect. As $\eta$ is increased further, additional secondary peaks and fluctuations are generated, which suppress quantum coherence. As the system approaches the adiabatic limit $\eta \gg \omega_i$ (right panel), coherence is restored and $P(\theta)$ is well described by a single Gaussian curve. In all depicted cases there is a finite photon pumping effect with $0.6 \leq \bar{n}_F \leq 1$. }
	\label{Fig:NAToA}
\end{figure*}
\appendix
\section{Appendix}\label{App:Appendix}

In this Appendix we provide further details on the frequency conversion effect. In Fig.~\ref{Fig:CVFreq} we plot the dependence of the frequency conversion efficiency $\bar{n}_F$ on $\eta_{\mbox{\tiny eff}}$, for a cavity with initially $N_{\mbox{\tiny ph}}=20$ photons in the topological class $B_0=1$, $B_m=8=B_d$, and four different commensurate frequency combinations of the form $\omega_1/\omega_2=q/p$. We note that as $q,p \gg1$, the nonadiabatic pumping effect vanishes, and the overall behavior resembles the one for irrationally-related frequencies.  

Moreover, we are interested in exploring whether the pumping effects persists for parameters outside the topological regime. We therefore plot $\bar{n}_F$ as a function of $\eta_{\mbox{\tiny eff}}$ for a cavity with initially $N_{\mbox{\tiny ph}}=20$ photons,  $B_0=1$, $B_d=8$ and $\omega_1/\omega_2=2/3$ and four different values of the Zeeman field amplitude. The topological phase boundary is at $B_m=12.5$. Although the pumping effects persists in the trivial phase of the model, the intensity diminishes as we go further away from the boundary. 
 
 Finally, for completeness we present the phase distribution $P(\theta)$ calculated in Sec.~\ref{sec:QuantCoher} for parameters ranging between the ultra-weak-drive up to the strong-drive limit. In Fig.~\ref{Fig:NAToA} we present $P(\theta)$ for seven values of $\eta$, $B_0=1$, $B_m=8$, $B_d=8$,  $N_{\mbox{\tiny ph}}=20$, and $\omega_1/\omega_2=2/3$. Phase coherence is preserved in the ultra-weak limit $\eta \ll \omega_i$, and $P(\theta)$ is described by a central Gaussian peak. As $\eta$ is increased, additional secondary peaks and phase fluctuations are generated and the pumped states is no longer characterized by a well defined phase. Phase coherence is restored in the strong-drive adiabatic limit, where $P(\theta)$ is a Gaussian curve at all times.

\pagebreak
\bibliography{FrequencyConversion}

\begin{thebibliography}{58}%
\makeatletter
\providecommand \@ifxundefined [1]{%
 \@ifx{#1\undefined}
}%
\providecommand \@ifnum [1]{%
 \ifnum #1\expandafter \@firstoftwo
 \else \expandafter \@secondoftwo
 \fi
}%
\providecommand \@ifx [1]{%
 \ifx #1\expandafter \@firstoftwo
 \else \expandafter \@secondoftwo
 \fi
}%
\providecommand \natexlab [1]{#1}%
\providecommand \enquote  [1]{``#1''}%
\providecommand \bibnamefont  [1]{#1}%
\providecommand \bibfnamefont [1]{#1}%
\providecommand \citenamefont [1]{#1}%
\providecommand \href@noop [0]{\@secondoftwo}%
\providecommand \href [0]{\begingroup \@sanitize@url \@href}%
\providecommand \@href[1]{\@@startlink{#1}\@@href}%
\providecommand \@@href[1]{\endgroup#1\@@endlink}%
\providecommand \@sanitize@url [0]{\catcode `\\12\catcode `\$12\catcode
  `\&12\catcode `\#12\catcode `\^12\catcode `\_12\catcode `\%12\relax}%
\providecommand \@@startlink[1]{}%
\providecommand \@@endlink[0]{}%
\providecommand \url  [0]{\begingroup\@sanitize@url \@url }%
\providecommand \@url [1]{\endgroup\@href {#1}{\urlprefix }}%
\providecommand \urlprefix  [0]{URL }%
\providecommand \Eprint [0]{\href }%
\providecommand \doibase [0]{http://dx.doi.org/}%
\providecommand \selectlanguage [0]{\@gobble}%
\providecommand \bibinfo  [0]{\@secondoftwo}%
\providecommand \bibfield  [0]{\@secondoftwo}%
\providecommand \translation [1]{[#1]}%
\providecommand \BibitemOpen [0]{}%
\providecommand \bibitemStop [0]{}%
\providecommand \bibitemNoStop [0]{.\EOS\space}%
\providecommand \EOS [0]{\spacefactor3000\relax}%
\providecommand \BibitemShut  [1]{\csname bibitem#1\endcsname}%
\let\auto@bib@innerbib\@empty
%</preamble>
\bibitem [{\citenamefont {Lauk}\ \emph {et~al.}(2020)\citenamefont {Lauk},
  \citenamefont {Sinclair}, \citenamefont {Barzanjeh}, \citenamefont {Covey},
  \citenamefont {Saffman}, \citenamefont {Spiropulu},\ and\ \citenamefont
  {Simon}}]{Lauk_2020}%
  \BibitemOpen
  \bibfield  {author} {\bibinfo {author} {\bibfnamefont {N.}~\bibnamefont
  {Lauk}}, \bibinfo {author} {\bibfnamefont {N.}~\bibnamefont {Sinclair}},
  \bibinfo {author} {\bibfnamefont {S.}~\bibnamefont {Barzanjeh}}, \bibinfo
  {author} {\bibfnamefont {J.~P.}\ \bibnamefont {Covey}}, \bibinfo {author}
  {\bibfnamefont {M.}~\bibnamefont {Saffman}}, \bibinfo {author} {\bibfnamefont
  {M.}~\bibnamefont {Spiropulu}}, \ and\ \bibinfo {author} {\bibfnamefont
  {C.}~\bibnamefont {Simon}},\ }\href {\doibase 10.1088/2058-9565/ab788a}
  {\bibfield  {journal} {\bibinfo  {journal} {Quantum Science and Technology}\
  }\textbf {\bibinfo {volume} {5}},\ \bibinfo {pages} {020501} (\bibinfo {year}
  {2020})}\BibitemShut {NoStop}%
\bibitem [{\citenamefont {O'Brien}\ \emph {et~al.}(2009)\citenamefont
  {O'Brien}, \citenamefont {Furusawa},\ and\ \citenamefont
  {Vu{\v{c}}kovi{\'{c}}}}]{O'Brien2009}%
  \BibitemOpen
  \bibfield  {author} {\bibinfo {author} {\bibfnamefont {J.~L.}\ \bibnamefont
  {O'Brien}}, \bibinfo {author} {\bibfnamefont {A.}~\bibnamefont {Furusawa}}, \
  and\ \bibinfo {author} {\bibfnamefont {J.}~\bibnamefont
  {Vu{\v{c}}kovi{\'{c}}}},\ }\href {\doibase 10.1038/nphoton.2009.229}
  {\bibfield  {journal} {\bibinfo  {journal} {Nature Photonics}\ }\textbf
  {\bibinfo {volume} {3}},\ \bibinfo {pages} {687} (\bibinfo {year}
  {2009})}\BibitemShut {NoStop}%
\bibitem [{\citenamefont {Briegel}\ \emph {et~al.}(1998)\citenamefont
  {Briegel}, \citenamefont {D\"ur}, \citenamefont {Cirac},\ and\ \citenamefont
  {Zoller}}]{PhysRevLett.81.5932}%
  \BibitemOpen
  \bibfield  {author} {\bibinfo {author} {\bibfnamefont {H.-J.}\ \bibnamefont
  {Briegel}}, \bibinfo {author} {\bibfnamefont {W.}~\bibnamefont {D\"ur}},
  \bibinfo {author} {\bibfnamefont {J.~I.}\ \bibnamefont {Cirac}}, \ and\
  \bibinfo {author} {\bibfnamefont {P.}~\bibnamefont {Zoller}},\ }\href
  {\doibase 10.1103/PhysRevLett.81.5932} {\bibfield  {journal} {\bibinfo
  {journal} {Phys. Rev. Lett.}\ }\textbf {\bibinfo {volume} {81}},\ \bibinfo
  {pages} {5932} (\bibinfo {year} {1998})}\BibitemShut {NoStop}%
\bibitem [{\citenamefont {Chaneli\`ere}\ \emph {et~al.}(2006)\citenamefont
  {Chaneli\`ere}, \citenamefont {Matsukevich}, \citenamefont {Jenkins},
  \citenamefont {Kennedy}, \citenamefont {Chapman},\ and\ \citenamefont
  {Kuzmich}}]{PhysRevLett.96.093604}%
  \BibitemOpen
  \bibfield  {author} {\bibinfo {author} {\bibfnamefont {T.}~\bibnamefont
  {Chaneli\`ere}}, \bibinfo {author} {\bibfnamefont {D.~N.}\ \bibnamefont
  {Matsukevich}}, \bibinfo {author} {\bibfnamefont {S.~D.}\ \bibnamefont
  {Jenkins}}, \bibinfo {author} {\bibfnamefont {T.~A.~B.}\ \bibnamefont
  {Kennedy}}, \bibinfo {author} {\bibfnamefont {M.~S.}\ \bibnamefont
  {Chapman}}, \ and\ \bibinfo {author} {\bibfnamefont {A.}~\bibnamefont
  {Kuzmich}},\ }\href {\doibase 10.1103/PhysRevLett.96.093604} {\bibfield
  {journal} {\bibinfo  {journal} {Phys. Rev. Lett.}\ }\textbf {\bibinfo
  {volume} {96}},\ \bibinfo {pages} {093604} (\bibinfo {year}
  {2006})}\BibitemShut {NoStop}%
\bibitem [{\citenamefont {Duan}\ \emph {et~al.}(2001)\citenamefont {Duan},
  \citenamefont {Lukin}, \citenamefont {Cirac},\ and\ \citenamefont
  {Zoller}}]{Duan2001}%
  \BibitemOpen
  \bibfield  {author} {\bibinfo {author} {\bibfnamefont {L.-M.}\ \bibnamefont
  {Duan}}, \bibinfo {author} {\bibfnamefont {M.~D.}\ \bibnamefont {Lukin}},
  \bibinfo {author} {\bibfnamefont {J.~I.}\ \bibnamefont {Cirac}}, \ and\
  \bibinfo {author} {\bibfnamefont {P.}~\bibnamefont {Zoller}},\ }\href
  {\doibase 10.1038/35106500} {\bibfield  {journal} {\bibinfo  {journal}
  {Nature}\ }\textbf {\bibinfo {volume} {414}},\ \bibinfo {pages} {413}
  (\bibinfo {year} {2001})}\BibitemShut {NoStop}%
\bibitem [{\citenamefont {Boozer}\ \emph {et~al.}(2007)\citenamefont {Boozer},
  \citenamefont {Boca}, \citenamefont {Miller}, \citenamefont {Northup},\ and\
  \citenamefont {Kimble}}]{PhysRevLett.98.193601}%
  \BibitemOpen
  \bibfield  {author} {\bibinfo {author} {\bibfnamefont {A.~D.}\ \bibnamefont
  {Boozer}}, \bibinfo {author} {\bibfnamefont {A.}~\bibnamefont {Boca}},
  \bibinfo {author} {\bibfnamefont {R.}~\bibnamefont {Miller}}, \bibinfo
  {author} {\bibfnamefont {T.~E.}\ \bibnamefont {Northup}}, \ and\ \bibinfo
  {author} {\bibfnamefont {H.~J.}\ \bibnamefont {Kimble}},\ }\href {\doibase
  10.1103/PhysRevLett.98.193601} {\bibfield  {journal} {\bibinfo  {journal}
  {Phys. Rev. Lett.}\ }\textbf {\bibinfo {volume} {98}},\ \bibinfo {pages}
  {193601} (\bibinfo {year} {2007})}\BibitemShut {NoStop}%
\bibitem [{\citenamefont {Chaneli{\`e}re}\ \emph {et~al.}(2005)\citenamefont
  {Chaneli{\`e}re}, \citenamefont {Matsukevich}, \citenamefont {Jenkins},
  \citenamefont {Lan}, \citenamefont {Kennedy},\ and\ \citenamefont
  {Kuzmich}}]{Chaneliere2005}%
  \BibitemOpen
  \bibfield  {author} {\bibinfo {author} {\bibfnamefont {T.}~\bibnamefont
  {Chaneli{\`e}re}}, \bibinfo {author} {\bibfnamefont {D.~N.}\ \bibnamefont
  {Matsukevich}}, \bibinfo {author} {\bibfnamefont {S.~D.}\ \bibnamefont
  {Jenkins}}, \bibinfo {author} {\bibfnamefont {S.-Y.}\ \bibnamefont {Lan}},
  \bibinfo {author} {\bibfnamefont {T.~A.~B.}\ \bibnamefont {Kennedy}}, \ and\
  \bibinfo {author} {\bibfnamefont {A.}~\bibnamefont {Kuzmich}},\ }\href
  {\doibase 10.1038/nature04315} {\bibfield  {journal} {\bibinfo  {journal}
  {Nature}\ }\textbf {\bibinfo {volume} {438}},\ \bibinfo {pages} {833}
  (\bibinfo {year} {2005})}\BibitemShut {NoStop}%
\bibitem [{\citenamefont {Olmschenk}\ \emph {et~al.}(2009)\citenamefont
  {Olmschenk}, \citenamefont {Matsukevich}, \citenamefont {Maunz},
  \citenamefont {Hayes}, \citenamefont {Duan},\ and\ \citenamefont
  {Monroe}}]{Olmschenk486}%
  \BibitemOpen
  \bibfield  {author} {\bibinfo {author} {\bibfnamefont {S.}~\bibnamefont
  {Olmschenk}}, \bibinfo {author} {\bibfnamefont {D.~N.}\ \bibnamefont
  {Matsukevich}}, \bibinfo {author} {\bibfnamefont {P.}~\bibnamefont {Maunz}},
  \bibinfo {author} {\bibfnamefont {D.}~\bibnamefont {Hayes}}, \bibinfo
  {author} {\bibfnamefont {L.-M.}\ \bibnamefont {Duan}}, \ and\ \bibinfo
  {author} {\bibfnamefont {C.}~\bibnamefont {Monroe}},\ }\href {\doibase
  10.1126/science.1167209} {\bibfield  {journal} {\bibinfo  {journal}
  {Science}\ }\textbf {\bibinfo {volume} {323}},\ \bibinfo {pages} {486}
  (\bibinfo {year} {2009})}\BibitemShut {NoStop}%
\bibitem [{\citenamefont {Gerardot}\ \emph {et~al.}(2008)\citenamefont
  {Gerardot}, \citenamefont {Brunner}, \citenamefont {Dalgarno}, \citenamefont
  {{\"O}hberg}, \citenamefont {Seidl}, \citenamefont {Kroner}, \citenamefont
  {Karrai}, \citenamefont {Stoltz}, \citenamefont {Petroff},\ and\
  \citenamefont {Warburton}}]{Gerardot2008}%
  \BibitemOpen
  \bibfield  {author} {\bibinfo {author} {\bibfnamefont {B.~D.}\ \bibnamefont
  {Gerardot}}, \bibinfo {author} {\bibfnamefont {D.}~\bibnamefont {Brunner}},
  \bibinfo {author} {\bibfnamefont {P.~A.}\ \bibnamefont {Dalgarno}}, \bibinfo
  {author} {\bibfnamefont {P.}~\bibnamefont {{\"O}hberg}}, \bibinfo {author}
  {\bibfnamefont {S.}~\bibnamefont {Seidl}}, \bibinfo {author} {\bibfnamefont
  {M.}~\bibnamefont {Kroner}}, \bibinfo {author} {\bibfnamefont
  {K.}~\bibnamefont {Karrai}}, \bibinfo {author} {\bibfnamefont {N.~G.}\
  \bibnamefont {Stoltz}}, \bibinfo {author} {\bibfnamefont {P.~M.}\
  \bibnamefont {Petroff}}, \ and\ \bibinfo {author} {\bibfnamefont {R.~J.}\
  \bibnamefont {Warburton}},\ }\href {\doibase 10.1038/nature06472} {\bibfield
  {journal} {\bibinfo  {journal} {Nature}\ }\textbf {\bibinfo {volume} {451}},\
  \bibinfo {pages} {441} (\bibinfo {year} {2008})}\BibitemShut {NoStop}%
\bibitem [{\citenamefont {Devoret}\ and\ \citenamefont
  {Schoelkopf}(2013)}]{Devoret1169}%
  \BibitemOpen
  \bibfield  {author} {\bibinfo {author} {\bibfnamefont {M.~H.}\ \bibnamefont
  {Devoret}}\ and\ \bibinfo {author} {\bibfnamefont {R.~J.}\ \bibnamefont
  {Schoelkopf}},\ }\href {\doibase 10.1126/science.1231930} {\bibfield
  {journal} {\bibinfo  {journal} {Science}\ }\textbf {\bibinfo {volume}
  {339}},\ \bibinfo {pages} {1169} (\bibinfo {year} {2013})}\BibitemShut
  {NoStop}%
\bibitem [{\citenamefont {Kimble}(2008)}]{Kimble2008}%
  \BibitemOpen
  \bibfield  {author} {\bibinfo {author} {\bibfnamefont {H.~J.}\ \bibnamefont
  {Kimble}},\ }\href {\doibase 10.1038/nature07127} {\bibfield  {journal}
  {\bibinfo  {journal} {Nature}\ }\textbf {\bibinfo {volume} {453}},\ \bibinfo
  {pages} {1023} (\bibinfo {year} {2008})}\BibitemShut {NoStop}%
\bibitem [{\citenamefont {Tanzilli}\ \emph {et~al.}(2005)\citenamefont
  {Tanzilli}, \citenamefont {Tittel}, \citenamefont {Halder}, \citenamefont
  {Alibart}, \citenamefont {Baldi}, \citenamefont {Gisin},\ and\ \citenamefont
  {Zbinden}}]{Tanzilli2005}%
  \BibitemOpen
  \bibfield  {author} {\bibinfo {author} {\bibfnamefont {S.}~\bibnamefont
  {Tanzilli}}, \bibinfo {author} {\bibfnamefont {W.}~\bibnamefont {Tittel}},
  \bibinfo {author} {\bibfnamefont {M.}~\bibnamefont {Halder}}, \bibinfo
  {author} {\bibfnamefont {O.}~\bibnamefont {Alibart}}, \bibinfo {author}
  {\bibfnamefont {P.}~\bibnamefont {Baldi}}, \bibinfo {author} {\bibfnamefont
  {N.}~\bibnamefont {Gisin}}, \ and\ \bibinfo {author} {\bibfnamefont
  {H.}~\bibnamefont {Zbinden}},\ }\href {\doibase 10.1038/nature04009}
  {\bibfield  {journal} {\bibinfo  {journal} {Nature}\ }\textbf {\bibinfo
  {volume} {437}},\ \bibinfo {pages} {116} (\bibinfo {year}
  {2005})}\BibitemShut {NoStop}%
\bibitem [{\citenamefont {Huang}\ and\ \citenamefont
  {Kumar}(1992)}]{PhysRevLett.68.2153}%
  \BibitemOpen
  \bibfield  {author} {\bibinfo {author} {\bibfnamefont {J.}~\bibnamefont
  {Huang}}\ and\ \bibinfo {author} {\bibfnamefont {P.}~\bibnamefont {Kumar}},\
  }\href {\doibase 10.1103/PhysRevLett.68.2153} {\bibfield  {journal} {\bibinfo
   {journal} {Phys. Rev. Lett.}\ }\textbf {\bibinfo {volume} {68}},\ \bibinfo
  {pages} {2153} (\bibinfo {year} {1992})}\BibitemShut {NoStop}%
\bibitem [{\citenamefont {Kumar}(1990)}]{Kumar:90}%
  \BibitemOpen
  \bibfield  {author} {\bibinfo {author} {\bibfnamefont {P.}~\bibnamefont
  {Kumar}},\ }\href {\doibase 10.1364/OL.15.001476} {\bibfield  {journal}
  {\bibinfo  {journal} {Opt. Lett.}\ }\textbf {\bibinfo {volume} {15}},\
  \bibinfo {pages} {1476} (\bibinfo {year} {1990})}\BibitemShut {NoStop}%
\bibitem [{\citenamefont {Mirhosseini}\ \emph {et~al.}(2020)\citenamefont
  {Mirhosseini}, \citenamefont {Sipahigil}, \citenamefont {Kalaee},\ and\
  \citenamefont {Painter}}]{Mirhosseini2020}%
  \BibitemOpen
  \bibfield  {author} {\bibinfo {author} {\bibfnamefont {M.}~\bibnamefont
  {Mirhosseini}}, \bibinfo {author} {\bibfnamefont {A.}~\bibnamefont
  {Sipahigil}}, \bibinfo {author} {\bibfnamefont {M.}~\bibnamefont {Kalaee}}, \
  and\ \bibinfo {author} {\bibfnamefont {O.}~\bibnamefont {Painter}},\ }\href
  {\doibase 10.1038/s41586-020-3038-6} {\bibfield  {journal} {\bibinfo
  {journal} {Nature}\ }\textbf {\bibinfo {volume} {588}},\ \bibinfo {pages}
  {599} (\bibinfo {year} {2020})}\BibitemShut {NoStop}%
\bibitem [{\citenamefont {Andrews}\ \emph {et~al.}(2014)\citenamefont
  {Andrews}, \citenamefont {Peterson}, \citenamefont {Purdy}, \citenamefont
  {Cicak}, \citenamefont {Simmonds}, \citenamefont {Regal},\ and\ \citenamefont
  {Lehnert}}]{Andrews2014}%
  \BibitemOpen
  \bibfield  {author} {\bibinfo {author} {\bibfnamefont {R.~W.}\ \bibnamefont
  {Andrews}}, \bibinfo {author} {\bibfnamefont {R.~W.}\ \bibnamefont
  {Peterson}}, \bibinfo {author} {\bibfnamefont {T.~P.}\ \bibnamefont {Purdy}},
  \bibinfo {author} {\bibfnamefont {K.}~\bibnamefont {Cicak}}, \bibinfo
  {author} {\bibfnamefont {R.~W.}\ \bibnamefont {Simmonds}}, \bibinfo {author}
  {\bibfnamefont {C.~A.}\ \bibnamefont {Regal}}, \ and\ \bibinfo {author}
  {\bibfnamefont {K.~W.}\ \bibnamefont {Lehnert}},\ }\href {\doibase
  10.1038/nphys2911} {\bibfield  {journal} {\bibinfo  {journal} {Nature
  Physics}\ }\textbf {\bibinfo {volume} {10}},\ \bibinfo {pages} {321}
  (\bibinfo {year} {2014})}\BibitemShut {NoStop}%
\bibitem [{\citenamefont {Rakher}\ \emph {et~al.}(2010)\citenamefont {Rakher},
  \citenamefont {Ma}, \citenamefont {Slattery}, \citenamefont {Tang},\ and\
  \citenamefont {Srinivasan}}]{Rakher2010}%
  \BibitemOpen
  \bibfield  {author} {\bibinfo {author} {\bibfnamefont {M.~T.}\ \bibnamefont
  {Rakher}}, \bibinfo {author} {\bibfnamefont {L.}~\bibnamefont {Ma}}, \bibinfo
  {author} {\bibfnamefont {O.}~\bibnamefont {Slattery}}, \bibinfo {author}
  {\bibfnamefont {X.}~\bibnamefont {Tang}}, \ and\ \bibinfo {author}
  {\bibfnamefont {K.}~\bibnamefont {Srinivasan}},\ }\href {\doibase
  10.1038/nphoton.2010.221} {\bibfield  {journal} {\bibinfo  {journal} {Nature
  Photonics}\ }\textbf {\bibinfo {volume} {4}},\ \bibinfo {pages} {786}
  (\bibinfo {year} {2010})}\BibitemShut {NoStop}%
\bibitem [{\citenamefont {Kitagawa}\ \emph {et~al.}(2010)\citenamefont
  {Kitagawa}, \citenamefont {Berg}, \citenamefont {Rudner},\ and\ \citenamefont
  {Demler}}]{PhysRevB.82.235114}%
  \BibitemOpen
  \bibfield  {author} {\bibinfo {author} {\bibfnamefont {T.}~\bibnamefont
  {Kitagawa}}, \bibinfo {author} {\bibfnamefont {E.}~\bibnamefont {Berg}},
  \bibinfo {author} {\bibfnamefont {M.}~\bibnamefont {Rudner}}, \ and\ \bibinfo
  {author} {\bibfnamefont {E.}~\bibnamefont {Demler}},\ }\href {\doibase
  10.1103/PhysRevB.82.235114} {\bibfield  {journal} {\bibinfo  {journal} {Phys.
  Rev. B}\ }\textbf {\bibinfo {volume} {82}},\ \bibinfo {pages} {235114}
  (\bibinfo {year} {2010})}\BibitemShut {NoStop}%
\bibitem [{\citenamefont {Lindner}\ \emph {et~al.}(2011)\citenamefont
  {Lindner}, \citenamefont {Refael},\ and\ \citenamefont
  {Galitski}}]{Lindner2011}%
  \BibitemOpen
  \bibfield  {author} {\bibinfo {author} {\bibfnamefont {N.~H.}\ \bibnamefont
  {Lindner}}, \bibinfo {author} {\bibfnamefont {G.}~\bibnamefont {Refael}}, \
  and\ \bibinfo {author} {\bibfnamefont {V.}~\bibnamefont {Galitski}},\ }\href
  {\doibase 10.1038/nphys1926} {\bibfield  {journal} {\bibinfo  {journal}
  {Nature Physics}\ }\textbf {\bibinfo {volume} {7}},\ \bibinfo {pages} {490}
  (\bibinfo {year} {2011})}\BibitemShut {NoStop}%
\bibitem [{\citenamefont {Potter}\ \emph {et~al.}(2016)\citenamefont {Potter},
  \citenamefont {Morimoto},\ and\ \citenamefont
  {Vishwanath}}]{PhysRevX.6.041001}%
  \BibitemOpen
  \bibfield  {author} {\bibinfo {author} {\bibfnamefont {A.~C.}\ \bibnamefont
  {Potter}}, \bibinfo {author} {\bibfnamefont {T.}~\bibnamefont {Morimoto}}, \
  and\ \bibinfo {author} {\bibfnamefont {A.}~\bibnamefont {Vishwanath}},\
  }\href {\doibase 10.1103/PhysRevX.6.041001} {\bibfield  {journal} {\bibinfo
  {journal} {Phys. Rev. X}\ }\textbf {\bibinfo {volume} {6}},\ \bibinfo {pages}
  {041001} (\bibinfo {year} {2016})}\BibitemShut {NoStop}%
\bibitem [{\citenamefont {Choi}\ \emph {et~al.}(2017)\citenamefont {Choi},
  \citenamefont {Choi}, \citenamefont {Landig}, \citenamefont {Kucsko},
  \citenamefont {Zhou}, \citenamefont {Isoya}, \citenamefont {Jelezko},
  \citenamefont {Onoda}, \citenamefont {Sumiya}, \citenamefont {Khemani},
  \citenamefont {von Keyserlingk}, \citenamefont {Yao}, \citenamefont
  {Demler},\ and\ \citenamefont {Lukin}}]{Choi2017}%
  \BibitemOpen
  \bibfield  {author} {\bibinfo {author} {\bibfnamefont {S.}~\bibnamefont
  {Choi}}, \bibinfo {author} {\bibfnamefont {J.}~\bibnamefont {Choi}}, \bibinfo
  {author} {\bibfnamefont {R.}~\bibnamefont {Landig}}, \bibinfo {author}
  {\bibfnamefont {G.}~\bibnamefont {Kucsko}}, \bibinfo {author} {\bibfnamefont
  {H.}~\bibnamefont {Zhou}}, \bibinfo {author} {\bibfnamefont {J.}~\bibnamefont
  {Isoya}}, \bibinfo {author} {\bibfnamefont {F.}~\bibnamefont {Jelezko}},
  \bibinfo {author} {\bibfnamefont {S.}~\bibnamefont {Onoda}}, \bibinfo
  {author} {\bibfnamefont {H.}~\bibnamefont {Sumiya}}, \bibinfo {author}
  {\bibfnamefont {V.}~\bibnamefont {Khemani}}, \bibinfo {author} {\bibfnamefont
  {C.}~\bibnamefont {von Keyserlingk}}, \bibinfo {author} {\bibfnamefont
  {N.~Y.}\ \bibnamefont {Yao}}, \bibinfo {author} {\bibfnamefont
  {E.}~\bibnamefont {Demler}}, \ and\ \bibinfo {author} {\bibfnamefont {M.~D.}\
  \bibnamefont {Lukin}},\ }\href {\doibase 10.1038/nature21426} {\bibfield
  {journal} {\bibinfo  {journal} {Nature}\ }\textbf {\bibinfo {volume} {543}},\
  \bibinfo {pages} {221} (\bibinfo {year} {2017})}\BibitemShut {NoStop}%
\bibitem [{\citenamefont {Peng}\ and\ \citenamefont
  {Refael}(2018)}]{PhysRevB.97.134303}%
  \BibitemOpen
  \bibfield  {author} {\bibinfo {author} {\bibfnamefont {Y.}~\bibnamefont
  {Peng}}\ and\ \bibinfo {author} {\bibfnamefont {G.}~\bibnamefont {Refael}},\
  }\href {\doibase 10.1103/PhysRevB.97.134303} {\bibfield  {journal} {\bibinfo
  {journal} {Phys. Rev. B}\ }\textbf {\bibinfo {volume} {97}},\ \bibinfo
  {pages} {134303} (\bibinfo {year} {2018})}\BibitemShut {NoStop}%
\bibitem [{\citenamefont {Martin}\ \emph {et~al.}(2017)\citenamefont {Martin},
  \citenamefont {Refael},\ and\ \citenamefont {Halperin}}]{PhysRevX.7.041008}%
  \BibitemOpen
  \bibfield  {author} {\bibinfo {author} {\bibfnamefont {I.}~\bibnamefont
  {Martin}}, \bibinfo {author} {\bibfnamefont {G.}~\bibnamefont {Refael}}, \
  and\ \bibinfo {author} {\bibfnamefont {B.}~\bibnamefont {Halperin}},\ }\href
  {\doibase 10.1103/PhysRevX.7.041008} {\bibfield  {journal} {\bibinfo
  {journal} {Phys. Rev. X}\ }\textbf {\bibinfo {volume} {7}},\ \bibinfo {pages}
  {041008} (\bibinfo {year} {2017})}\BibitemShut {NoStop}%
\bibitem [{\citenamefont {Crowley}\ \emph {et~al.}(2019)\citenamefont
  {Crowley}, \citenamefont {Martin},\ and\ \citenamefont
  {Chandran}}]{PhysRevB.99.064306}%
  \BibitemOpen
  \bibfield  {author} {\bibinfo {author} {\bibfnamefont {P.~J.~D.}\
  \bibnamefont {Crowley}}, \bibinfo {author} {\bibfnamefont {I.}~\bibnamefont
  {Martin}}, \ and\ \bibinfo {author} {\bibfnamefont {A.}~\bibnamefont
  {Chandran}},\ }\href {\doibase 10.1103/PhysRevB.99.064306} {\bibfield
  {journal} {\bibinfo  {journal} {Phys. Rev. B}\ }\textbf {\bibinfo {volume}
  {99}},\ \bibinfo {pages} {064306} (\bibinfo {year} {2019})}\BibitemShut
  {NoStop}%
\bibitem [{\citenamefont {Malz}\ and\ \citenamefont
  {Smith}(2020)}]{2012.01459}%
  \BibitemOpen
  \bibfield  {author} {\bibinfo {author} {\bibfnamefont {D.}~\bibnamefont
  {Malz}}\ and\ \bibinfo {author} {\bibfnamefont {A.}~\bibnamefont {Smith}},\
  }\href@noop {} {\enquote {\bibinfo {title} {Topological two-dimensional
  floquet lattice on a single superconducting qubit},}\ } (\bibinfo {year}
  {2020}),\ \Eprint {http://arxiv.org/abs/arXiv:2012.01459} {arXiv:2012.01459}
  \BibitemShut {NoStop}%
\bibitem [{\citenamefont {Nathan}\ \emph {et~al.}(2019)\citenamefont {Nathan},
  \citenamefont {Martin},\ and\ \citenamefont {Refael}}]{PhysRevB.99.094311}%
  \BibitemOpen
  \bibfield  {author} {\bibinfo {author} {\bibfnamefont {F.}~\bibnamefont
  {Nathan}}, \bibinfo {author} {\bibfnamefont {I.}~\bibnamefont {Martin}}, \
  and\ \bibinfo {author} {\bibfnamefont {G.}~\bibnamefont {Refael}},\ }\href
  {\doibase 10.1103/PhysRevB.99.094311} {\bibfield  {journal} {\bibinfo
  {journal} {Phys. Rev. B}\ }\textbf {\bibinfo {volume} {99}},\ \bibinfo
  {pages} {094311} (\bibinfo {year} {2019})}\BibitemShut {NoStop}%
\bibitem [{\citenamefont {Abe}\ \emph {et~al.}(2011)\citenamefont {Abe},
  \citenamefont {Wu}, \citenamefont {Ardavan},\ and\ \citenamefont
  {Morton}}]{doi:10.1063/1.3601930}%
  \BibitemOpen
  \bibfield  {author} {\bibinfo {author} {\bibfnamefont {E.}~\bibnamefont
  {Abe}}, \bibinfo {author} {\bibfnamefont {H.}~\bibnamefont {Wu}}, \bibinfo
  {author} {\bibfnamefont {A.}~\bibnamefont {Ardavan}}, \ and\ \bibinfo
  {author} {\bibfnamefont {J.~J.~L.}\ \bibnamefont {Morton}},\ }\href {\doibase
  10.1063/1.3601930} {\bibfield  {journal} {\bibinfo  {journal} {Applied
  Physics Letters}\ }\textbf {\bibinfo {volume} {98}},\ \bibinfo {pages}
  {251108} (\bibinfo {year} {2011})}\BibitemShut {NoStop}%
\bibitem [{\citenamefont {Anderson}(1958)}]{PhysRev.109.1492}%
  \BibitemOpen
  \bibfield  {author} {\bibinfo {author} {\bibfnamefont {P.~W.}\ \bibnamefont
  {Anderson}},\ }\href {\doibase 10.1103/PhysRev.109.1492} {\bibfield
  {journal} {\bibinfo  {journal} {Phys. Rev.}\ }\textbf {\bibinfo {volume}
  {109}},\ \bibinfo {pages} {1492} (\bibinfo {year} {1958})}\BibitemShut
  {NoStop}%
\bibitem [{\citenamefont {Wegner}(1980)}]{Wegner1980}%
  \BibitemOpen
  \bibfield  {author} {\bibinfo {author} {\bibfnamefont {F.}~\bibnamefont
  {Wegner}},\ }\href {\doibase 10.1007/BF01325284} {\bibfield  {journal}
  {\bibinfo  {journal} {Zeitschrift f{\"u}r Physik B Condensed Matter}\
  }\textbf {\bibinfo {volume} {36}},\ \bibinfo {pages} {209} (\bibinfo {year}
  {1980})}\BibitemShut {NoStop}%
\bibitem [{\citenamefont {Thouless}(1983)}]{PhysRevB.27.6083}%
  \BibitemOpen
  \bibfield  {author} {\bibinfo {author} {\bibfnamefont {D.~J.}\ \bibnamefont
  {Thouless}},\ }\href {\doibase 10.1103/PhysRevB.27.6083} {\bibfield
  {journal} {\bibinfo  {journal} {Phys. Rev. B}\ }\textbf {\bibinfo {volume}
  {27}},\ \bibinfo {pages} {6083} (\bibinfo {year} {1983})}\BibitemShut
  {NoStop}%
\bibitem [{\citenamefont {Nathan}\ \emph {et~al.}(2020)\citenamefont {Nathan},
  \citenamefont {Refael}, \citenamefont {Rudner},\ and\ \citenamefont
  {Martin}}]{PhysRev.Research.2.043411}%
  \BibitemOpen
  \bibfield  {author} {\bibinfo {author} {\bibfnamefont {F.}~\bibnamefont
  {Nathan}}, \bibinfo {author} {\bibfnamefont {G.}~\bibnamefont {Refael}},
  \bibinfo {author} {\bibfnamefont {M.~S.}\ \bibnamefont {Rudner}}, \ and\
  \bibinfo {author} {\bibfnamefont {I.}~\bibnamefont {Martin}},\ }\href
  {\doibase 10.1103/physrevresearch.2.043411} {\bibfield  {journal} {\bibinfo
  {journal} {Physical Review Research}\ }\textbf {\bibinfo {volume} {2}}
  (\bibinfo {year} {2020}),\ 10.1103/physrevresearch.2.043411}\BibitemShut
  {NoStop}%
\bibitem [{\citenamefont {Weinberg}\ and\ \citenamefont
  {Bukov}(2017)}]{SciPostPhys.2.1.003}%
  \BibitemOpen
  \bibfield  {author} {\bibinfo {author} {\bibfnamefont {P.}~\bibnamefont
  {Weinberg}}\ and\ \bibinfo {author} {\bibfnamefont {M.}~\bibnamefont
  {Bukov}},\ }\href {\doibase 10.21468/SciPostPhys.2.1.003} {\bibfield
  {journal} {\bibinfo  {journal} {SciPost Phys.}\ }\textbf {\bibinfo {volume}
  {2}},\ \bibinfo {pages} {003} (\bibinfo {year} {2017})}\BibitemShut {NoStop}%
\bibitem [{\citenamefont {Privitera}\ \emph {et~al.}(2018)\citenamefont
  {Privitera}, \citenamefont {Russomanno}, \citenamefont {Citro},\ and\
  \citenamefont {Santoro}}]{PhysRevLett.120.106601}%
  \BibitemOpen
  \bibfield  {author} {\bibinfo {author} {\bibfnamefont {L.}~\bibnamefont
  {Privitera}}, \bibinfo {author} {\bibfnamefont {A.}~\bibnamefont
  {Russomanno}}, \bibinfo {author} {\bibfnamefont {R.}~\bibnamefont {Citro}}, \
  and\ \bibinfo {author} {\bibfnamefont {G.~E.}\ \bibnamefont {Santoro}},\
  }\href {\doibase 10.1103/PhysRevLett.120.106601} {\bibfield  {journal}
  {\bibinfo  {journal} {Phys. Rev. Lett.}\ }\textbf {\bibinfo {volume} {120}},\
  \bibinfo {pages} {106601} (\bibinfo {year} {2018})}\BibitemShut {NoStop}%
\bibitem [{\citenamefont {Psaroudaki}\ and\ \citenamefont
  {Refael}(2021)}]{Psaroudaki2021}%
  \BibitemOpen
  \bibfield  {author} {\bibinfo {author} {\bibfnamefont {C.}~\bibnamefont
  {Psaroudaki}}\ and\ \bibinfo {author} {\bibfnamefont {G.}~\bibnamefont
  {Refael}},\ }\href@noop {} {\enquote {\bibinfo {title} {Energy transfer in a
  random-matrix floquet hamiltonian},}\ } (\bibinfo {year} {2021}),\ \Eprint
  {http://arxiv.org/abs/manuscript in prepration} {manuscript in prepration}
  \BibitemShut {NoStop}%
\bibitem [{\citenamefont {Shirley}(1965)}]{PhysRev.138.B979}%
  \BibitemOpen
  \bibfield  {author} {\bibinfo {author} {\bibfnamefont {J.~H.}\ \bibnamefont
  {Shirley}},\ }\href {\doibase 10.1103/PhysRev.138.B979} {\bibfield  {journal}
  {\bibinfo  {journal} {Phys. Rev.}\ }\textbf {\bibinfo {volume} {138}},\
  \bibinfo {pages} {B979} (\bibinfo {year} {1965})}\BibitemShut {NoStop}%
\bibitem [{\citenamefont {Romito}\ \emph {et~al.}(2018)\citenamefont {Romito},
  \citenamefont {Lobo},\ and\ \citenamefont {Recati}}]{Romito2018}%
  \BibitemOpen
  \bibfield  {author} {\bibinfo {author} {\bibfnamefont {D.}~\bibnamefont
  {Romito}}, \bibinfo {author} {\bibfnamefont {C.}~\bibnamefont {Lobo}}, \ and\
  \bibinfo {author} {\bibfnamefont {A.}~\bibnamefont {Recati}},\ }\href
  {\doibase 10.1140/epjd/e2018-90081-3} {\bibfield  {journal} {\bibinfo
  {journal} {The European Physical Journal D}\ }\textbf {\bibinfo {volume}
  {72}},\ \bibinfo {pages} {135} (\bibinfo {year} {2018})}\BibitemShut
  {NoStop}%
\bibitem [{\citenamefont {Hofstadter}(1976)}]{PhysRevB.14.2239}%
  \BibitemOpen
  \bibfield  {author} {\bibinfo {author} {\bibfnamefont {D.~R.}\ \bibnamefont
  {Hofstadter}},\ }\href {\doibase 10.1103/PhysRevB.14.2239} {\bibfield
  {journal} {\bibinfo  {journal} {Phys. Rev. B}\ }\textbf {\bibinfo {volume}
  {14}},\ \bibinfo {pages} {2239} (\bibinfo {year} {1976})}\BibitemShut
  {NoStop}%
\bibitem [{\citenamefont {Evers}\ and\ \citenamefont
  {Mirlin}(2008)}]{RevModPhys.80.1355}%
  \BibitemOpen
  \bibfield  {author} {\bibinfo {author} {\bibfnamefont {F.}~\bibnamefont
  {Evers}}\ and\ \bibinfo {author} {\bibfnamefont {A.~D.}\ \bibnamefont
  {Mirlin}},\ }\href {\doibase 10.1103/RevModPhys.80.1355} {\bibfield
  {journal} {\bibinfo  {journal} {Rev. Mod. Phys.}\ }\textbf {\bibinfo {volume}
  {80}},\ \bibinfo {pages} {1355} (\bibinfo {year} {2008})}\BibitemShut
  {NoStop}%
\bibitem [{\citenamefont {Mirlin}\ \emph {et~al.}(1996)\citenamefont {Mirlin},
  \citenamefont {Fyodorov}, \citenamefont {Dittes}, \citenamefont {Quezada},\
  and\ \citenamefont {Seligman}}]{PhysRevE.54.3221}%
  \BibitemOpen
  \bibfield  {author} {\bibinfo {author} {\bibfnamefont {A.~D.}\ \bibnamefont
  {Mirlin}}, \bibinfo {author} {\bibfnamefont {Y.~V.}\ \bibnamefont
  {Fyodorov}}, \bibinfo {author} {\bibfnamefont {F.-M.}\ \bibnamefont
  {Dittes}}, \bibinfo {author} {\bibfnamefont {J.}~\bibnamefont {Quezada}}, \
  and\ \bibinfo {author} {\bibfnamefont {T.~H.}\ \bibnamefont {Seligman}},\
  }\href {\doibase 10.1103/PhysRevE.54.3221} {\bibfield  {journal} {\bibinfo
  {journal} {Phys. Rev. E}\ }\textbf {\bibinfo {volume} {54}},\ \bibinfo
  {pages} {3221} (\bibinfo {year} {1996})}\BibitemShut {NoStop}%
\bibitem [{\citenamefont {Wang}\ and\ \citenamefont
  {Gong}(2009)}]{PhysRevLett.102.244102}%
  \BibitemOpen
  \bibfield  {author} {\bibinfo {author} {\bibfnamefont {J.}~\bibnamefont
  {Wang}}\ and\ \bibinfo {author} {\bibfnamefont {J.}~\bibnamefont {Gong}},\
  }\href {\doibase 10.1103/PhysRevLett.102.244102} {\bibfield  {journal}
  {\bibinfo  {journal} {Phys. Rev. Lett.}\ }\textbf {\bibinfo {volume} {102}},\
  \bibinfo {pages} {244102} (\bibinfo {year} {2009})}\BibitemShut {NoStop}%
\bibitem [{\citenamefont {Pegg}\ and\ \citenamefont
  {Barnett}(1989)}]{PhysRevA.39.1665}%
  \BibitemOpen
  \bibfield  {author} {\bibinfo {author} {\bibfnamefont {D.~T.}\ \bibnamefont
  {Pegg}}\ and\ \bibinfo {author} {\bibfnamefont {S.~M.}\ \bibnamefont
  {Barnett}},\ }\href {\doibase 10.1103/PhysRevA.39.1665} {\bibfield  {journal}
  {\bibinfo  {journal} {Phys. Rev. A}\ }\textbf {\bibinfo {volume} {39}},\
  \bibinfo {pages} {1665} (\bibinfo {year} {1989})}\BibitemShut {NoStop}%
\bibitem [{\citenamefont {Walls}\ and\ \citenamefont
  {Milburn}(1995)}]{walls_milburn_1995}%
  \BibitemOpen
  \bibfield  {author} {\bibinfo {author} {\bibfnamefont {D.~F.}\ \bibnamefont
  {Walls}}\ and\ \bibinfo {author} {\bibfnamefont {G.~J.}\ \bibnamefont
  {Milburn}},\ }\href@noop {} {\emph {\bibinfo {title} {Quantum optics}}}\
  (\bibinfo  {publisher} {Springer},\ \bibinfo {year} {1995})\BibitemShut
  {NoStop}%
\bibitem [{\citenamefont {Tavis}\ and\ \citenamefont
  {Cummings}(1968)}]{PhysRev.170.379}%
  \BibitemOpen
  \bibfield  {author} {\bibinfo {author} {\bibfnamefont {M.}~\bibnamefont
  {Tavis}}\ and\ \bibinfo {author} {\bibfnamefont {F.~W.}\ \bibnamefont
  {Cummings}},\ }\href {\doibase 10.1103/PhysRev.170.379} {\bibfield  {journal}
  {\bibinfo  {journal} {Phys. Rev.}\ }\textbf {\bibinfo {volume} {170}},\
  \bibinfo {pages} {379} (\bibinfo {year} {1968})}\BibitemShut {NoStop}%
\bibitem [{\citenamefont {Huebl}\ \emph {et~al.}(2013)\citenamefont {Huebl},
  \citenamefont {Zollitsch}, \citenamefont {Lotze}, \citenamefont {Hocke},
  \citenamefont {Greifenstein}, \citenamefont {Marx}, \citenamefont {Gross},\
  and\ \citenamefont {Goennenwein}}]{PhysRevLett.111.127003}%
  \BibitemOpen
  \bibfield  {author} {\bibinfo {author} {\bibfnamefont {H.}~\bibnamefont
  {Huebl}}, \bibinfo {author} {\bibfnamefont {C.~W.}\ \bibnamefont
  {Zollitsch}}, \bibinfo {author} {\bibfnamefont {J.}~\bibnamefont {Lotze}},
  \bibinfo {author} {\bibfnamefont {F.}~\bibnamefont {Hocke}}, \bibinfo
  {author} {\bibfnamefont {M.}~\bibnamefont {Greifenstein}}, \bibinfo {author}
  {\bibfnamefont {A.}~\bibnamefont {Marx}}, \bibinfo {author} {\bibfnamefont
  {R.}~\bibnamefont {Gross}}, \ and\ \bibinfo {author} {\bibfnamefont
  {S.~T.~B.}\ \bibnamefont {Goennenwein}},\ }\href {\doibase
  10.1103/PhysRevLett.111.127003} {\bibfield  {journal} {\bibinfo  {journal}
  {Phys. Rev. Lett.}\ }\textbf {\bibinfo {volume} {111}},\ \bibinfo {pages}
  {127003} (\bibinfo {year} {2013})}\BibitemShut {NoStop}%
\bibitem [{\citenamefont {Tabuchi}\ \emph {et~al.}(2014)\citenamefont
  {Tabuchi}, \citenamefont {Ishino}, \citenamefont {Ishikawa}, \citenamefont
  {Yamazaki}, \citenamefont {Usami},\ and\ \citenamefont
  {Nakamura}}]{PhysRevLett.113.083603}%
  \BibitemOpen
  \bibfield  {author} {\bibinfo {author} {\bibfnamefont {Y.}~\bibnamefont
  {Tabuchi}}, \bibinfo {author} {\bibfnamefont {S.}~\bibnamefont {Ishino}},
  \bibinfo {author} {\bibfnamefont {T.}~\bibnamefont {Ishikawa}}, \bibinfo
  {author} {\bibfnamefont {R.}~\bibnamefont {Yamazaki}}, \bibinfo {author}
  {\bibfnamefont {K.}~\bibnamefont {Usami}}, \ and\ \bibinfo {author}
  {\bibfnamefont {Y.}~\bibnamefont {Nakamura}},\ }\href {\doibase
  10.1103/PhysRevLett.113.083603} {\bibfield  {journal} {\bibinfo  {journal}
  {Phys. Rev. Lett.}\ }\textbf {\bibinfo {volume} {113}},\ \bibinfo {pages}
  {083603} (\bibinfo {year} {2014})}\BibitemShut {NoStop}%
\bibitem [{\citenamefont {Zhang}\ \emph {et~al.}(2014)\citenamefont {Zhang},
  \citenamefont {Zou}, \citenamefont {Jiang},\ and\ \citenamefont
  {Tang}}]{PhysRevLett.113.156401}%
  \BibitemOpen
  \bibfield  {author} {\bibinfo {author} {\bibfnamefont {X.}~\bibnamefont
  {Zhang}}, \bibinfo {author} {\bibfnamefont {C.-L.}\ \bibnamefont {Zou}},
  \bibinfo {author} {\bibfnamefont {L.}~\bibnamefont {Jiang}}, \ and\ \bibinfo
  {author} {\bibfnamefont {H.~X.}\ \bibnamefont {Tang}},\ }\href {\doibase
  10.1103/PhysRevLett.113.156401} {\bibfield  {journal} {\bibinfo  {journal}
  {Phys. Rev. Lett.}\ }\textbf {\bibinfo {volume} {113}},\ \bibinfo {pages}
  {156401} (\bibinfo {year} {2014})}\BibitemShut {NoStop}%
\bibitem [{\citenamefont {Maier-Flaig}\ \emph {et~al.}(2016)\citenamefont
  {Maier-Flaig}, \citenamefont {Harder}, \citenamefont {Gross}, \citenamefont
  {Huebl},\ and\ \citenamefont {Goennenwein}}]{PhysRevB.94.054433}%
  \BibitemOpen
  \bibfield  {author} {\bibinfo {author} {\bibfnamefont {H.}~\bibnamefont
  {Maier-Flaig}}, \bibinfo {author} {\bibfnamefont {M.}~\bibnamefont {Harder}},
  \bibinfo {author} {\bibfnamefont {R.}~\bibnamefont {Gross}}, \bibinfo
  {author} {\bibfnamefont {H.}~\bibnamefont {Huebl}}, \ and\ \bibinfo {author}
  {\bibfnamefont {S.~T.~B.}\ \bibnamefont {Goennenwein}},\ }\href {\doibase
  10.1103/PhysRevB.94.054433} {\bibfield  {journal} {\bibinfo  {journal} {Phys.
  Rev. B}\ }\textbf {\bibinfo {volume} {94}},\ \bibinfo {pages} {054433}
  (\bibinfo {year} {2016})}\BibitemShut {NoStop}%
\bibitem [{\citenamefont {Kostylev}\ \emph {et~al.}(2016)\citenamefont
  {Kostylev}, \citenamefont {Goryachev},\ and\ \citenamefont
  {Tobar}}]{doi:10.1063/1.4941730}%
  \BibitemOpen
  \bibfield  {author} {\bibinfo {author} {\bibfnamefont {N.}~\bibnamefont
  {Kostylev}}, \bibinfo {author} {\bibfnamefont {M.}~\bibnamefont {Goryachev}},
  \ and\ \bibinfo {author} {\bibfnamefont {M.~E.}\ \bibnamefont {Tobar}},\
  }\href {\doibase 10.1063/1.4941730} {\bibfield  {journal} {\bibinfo
  {journal} {Applied Physics Letters}\ }\textbf {\bibinfo {volume} {108}},\
  \bibinfo {pages} {062402} (\bibinfo {year} {2016})}\BibitemShut {NoStop}%
\bibitem [{\citenamefont {Kubo}\ \emph {et~al.}(2010)\citenamefont {Kubo},
  \citenamefont {Ong}, \citenamefont {Bertet}, \citenamefont {Vion},
  \citenamefont {Jacques}, \citenamefont {Zheng}, \citenamefont {Dr\'eau},
  \citenamefont {Roch}, \citenamefont {Auffeves}, \citenamefont {Jelezko},
  \citenamefont {Wrachtrup}, \citenamefont {Barthe}, \citenamefont {Bergonzo},\
  and\ \citenamefont {Esteve}}]{PhysRevLett.105.140502}%
  \BibitemOpen
  \bibfield  {author} {\bibinfo {author} {\bibfnamefont {Y.}~\bibnamefont
  {Kubo}}, \bibinfo {author} {\bibfnamefont {F.~R.}\ \bibnamefont {Ong}},
  \bibinfo {author} {\bibfnamefont {P.}~\bibnamefont {Bertet}}, \bibinfo
  {author} {\bibfnamefont {D.}~\bibnamefont {Vion}}, \bibinfo {author}
  {\bibfnamefont {V.}~\bibnamefont {Jacques}}, \bibinfo {author} {\bibfnamefont
  {D.}~\bibnamefont {Zheng}}, \bibinfo {author} {\bibfnamefont
  {A.}~\bibnamefont {Dr\'eau}}, \bibinfo {author} {\bibfnamefont {J.-F.}\
  \bibnamefont {Roch}}, \bibinfo {author} {\bibfnamefont {A.}~\bibnamefont
  {Auffeves}}, \bibinfo {author} {\bibfnamefont {F.}~\bibnamefont {Jelezko}},
  \bibinfo {author} {\bibfnamefont {J.}~\bibnamefont {Wrachtrup}}, \bibinfo
  {author} {\bibfnamefont {M.~F.}\ \bibnamefont {Barthe}}, \bibinfo {author}
  {\bibfnamefont {P.}~\bibnamefont {Bergonzo}}, \ and\ \bibinfo {author}
  {\bibfnamefont {D.}~\bibnamefont {Esteve}},\ }\href {\doibase
  10.1103/PhysRevLett.105.140502} {\bibfield  {journal} {\bibinfo  {journal}
  {Phys. Rev. Lett.}\ }\textbf {\bibinfo {volume} {105}},\ \bibinfo {pages}
  {140502} (\bibinfo {year} {2010})}\BibitemShut {NoStop}%
\bibitem [{\citenamefont {Schuster}\ \emph {et~al.}(2010)\citenamefont
  {Schuster}, \citenamefont {Sears}, \citenamefont {Ginossar}, \citenamefont
  {DiCarlo}, \citenamefont {Frunzio}, \citenamefont {Morton}, \citenamefont
  {Wu}, \citenamefont {Briggs}, \citenamefont {Buckley}, \citenamefont
  {Awschalom},\ and\ \citenamefont {Schoelkopf}}]{PhysRevLett.105.140501}%
  \BibitemOpen
  \bibfield  {author} {\bibinfo {author} {\bibfnamefont {D.~I.}\ \bibnamefont
  {Schuster}}, \bibinfo {author} {\bibfnamefont {A.~P.}\ \bibnamefont {Sears}},
  \bibinfo {author} {\bibfnamefont {E.}~\bibnamefont {Ginossar}}, \bibinfo
  {author} {\bibfnamefont {L.}~\bibnamefont {DiCarlo}}, \bibinfo {author}
  {\bibfnamefont {L.}~\bibnamefont {Frunzio}}, \bibinfo {author} {\bibfnamefont
  {J.~J.~L.}\ \bibnamefont {Morton}}, \bibinfo {author} {\bibfnamefont
  {H.}~\bibnamefont {Wu}}, \bibinfo {author} {\bibfnamefont {G.~A.~D.}\
  \bibnamefont {Briggs}}, \bibinfo {author} {\bibfnamefont {B.~B.}\
  \bibnamefont {Buckley}}, \bibinfo {author} {\bibfnamefont {D.~D.}\
  \bibnamefont {Awschalom}}, \ and\ \bibinfo {author} {\bibfnamefont {R.~J.}\
  \bibnamefont {Schoelkopf}},\ }\href {\doibase 10.1103/PhysRevLett.105.140501}
  {\bibfield  {journal} {\bibinfo  {journal} {Phys. Rev. Lett.}\ }\textbf
  {\bibinfo {volume} {105}},\ \bibinfo {pages} {140501} (\bibinfo {year}
  {2010})}\BibitemShut {NoStop}%
\bibitem [{\citenamefont {Gimeno}\ \emph {et~al.}(2020)\citenamefont {Gimeno},
  \citenamefont {Kersten}, \citenamefont {Pallar{\'e}s}, \citenamefont
  {Hermosilla}, \citenamefont {Mart{\'i}nez-P{\'e}rez}, \citenamefont
  {Jenkins}, \citenamefont {Angerer}, \citenamefont {S{\'a}nchez-Azqueta},
  \citenamefont {Zueco}, \citenamefont {Majer}, \citenamefont {Lostao},\ and\
  \citenamefont {Luis}}]{Gimeno2020}%
  \BibitemOpen
  \bibfield  {author} {\bibinfo {author} {\bibfnamefont {I.}~\bibnamefont
  {Gimeno}}, \bibinfo {author} {\bibfnamefont {W.}~\bibnamefont {Kersten}},
  \bibinfo {author} {\bibfnamefont {M.~C.}\ \bibnamefont {Pallar{\'e}s}},
  \bibinfo {author} {\bibfnamefont {P.}~\bibnamefont {Hermosilla}}, \bibinfo
  {author} {\bibfnamefont {M.~J.}\ \bibnamefont {Mart{\'i}nez-P{\'e}rez}},
  \bibinfo {author} {\bibfnamefont {M.~D.}\ \bibnamefont {Jenkins}}, \bibinfo
  {author} {\bibfnamefont {A.}~\bibnamefont {Angerer}}, \bibinfo {author}
  {\bibfnamefont {C.}~\bibnamefont {S{\'a}nchez-Azqueta}}, \bibinfo {author}
  {\bibfnamefont {D.}~\bibnamefont {Zueco}}, \bibinfo {author} {\bibfnamefont
  {J.}~\bibnamefont {Majer}}, \bibinfo {author} {\bibfnamefont
  {A.}~\bibnamefont {Lostao}}, \ and\ \bibinfo {author} {\bibfnamefont
  {F.}~\bibnamefont {Luis}},\ }\href {\doibase 10.1021/acsnano.0c03167}
  {\bibfield  {journal} {\bibinfo  {journal} {ACS Nano}\ }\textbf {\bibinfo
  {volume} {14}},\ \bibinfo {pages} {8707} (\bibinfo {year}
  {2020})}\BibitemShut {NoStop}%
\bibitem [{\citenamefont {K\"orber}\ \emph {et~al.}(2020)\citenamefont
  {K\"orber}, \citenamefont {Privitera}, \citenamefont {Budich},\ and\
  \citenamefont {Trauzettel}}]{PhysRevResearch.2.022023}%
  \BibitemOpen
  \bibfield  {author} {\bibinfo {author} {\bibfnamefont {S.}~\bibnamefont
  {K\"orber}}, \bibinfo {author} {\bibfnamefont {L.}~\bibnamefont {Privitera}},
  \bibinfo {author} {\bibfnamefont {J.~C.}\ \bibnamefont {Budich}}, \ and\
  \bibinfo {author} {\bibfnamefont {B.}~\bibnamefont {Trauzettel}},\ }\href
  {\doibase 10.1103/PhysRevResearch.2.022023} {\bibfield  {journal} {\bibinfo
  {journal} {Phys. Rev. Research}\ }\textbf {\bibinfo {volume} {2}},\ \bibinfo
  {pages} {022023} (\bibinfo {year} {2020})}\BibitemShut {NoStop}%
\bibitem [{\citenamefont {Fink}\ \emph {et~al.}(2009)\citenamefont {Fink},
  \citenamefont {Bianchetti}, \citenamefont {Baur}, \citenamefont {G\"oppl},
  \citenamefont {Steffen}, \citenamefont {Filipp}, \citenamefont {Leek},
  \citenamefont {Blais},\ and\ \citenamefont
  {Wallraff}}]{PhysRevLett.103.083601}%
  \BibitemOpen
  \bibfield  {author} {\bibinfo {author} {\bibfnamefont {J.~M.}\ \bibnamefont
  {Fink}}, \bibinfo {author} {\bibfnamefont {R.}~\bibnamefont {Bianchetti}},
  \bibinfo {author} {\bibfnamefont {M.}~\bibnamefont {Baur}}, \bibinfo {author}
  {\bibfnamefont {M.}~\bibnamefont {G\"oppl}}, \bibinfo {author} {\bibfnamefont
  {L.}~\bibnamefont {Steffen}}, \bibinfo {author} {\bibfnamefont
  {S.}~\bibnamefont {Filipp}}, \bibinfo {author} {\bibfnamefont {P.~J.}\
  \bibnamefont {Leek}}, \bibinfo {author} {\bibfnamefont {A.}~\bibnamefont
  {Blais}}, \ and\ \bibinfo {author} {\bibfnamefont {A.}~\bibnamefont
  {Wallraff}},\ }\href {\doibase 10.1103/PhysRevLett.103.083601} {\bibfield
  {journal} {\bibinfo  {journal} {Phys. Rev. Lett.}\ }\textbf {\bibinfo
  {volume} {103}},\ \bibinfo {pages} {083601} (\bibinfo {year}
  {2009})}\BibitemShut {NoStop}%
\bibitem [{\citenamefont {Mi}\ \emph {et~al.}(2017)\citenamefont {Mi},
  \citenamefont {Cady}, \citenamefont {Zajac}, \citenamefont {Deelman},\ and\
  \citenamefont {Petta}}]{Mi156}%
  \BibitemOpen
  \bibfield  {author} {\bibinfo {author} {\bibfnamefont {X.}~\bibnamefont
  {Mi}}, \bibinfo {author} {\bibfnamefont {J.~V.}\ \bibnamefont {Cady}},
  \bibinfo {author} {\bibfnamefont {D.~M.}\ \bibnamefont {Zajac}}, \bibinfo
  {author} {\bibfnamefont {P.~W.}\ \bibnamefont {Deelman}}, \ and\ \bibinfo
  {author} {\bibfnamefont {J.~R.}\ \bibnamefont {Petta}},\ }\href {\doibase
  10.1126/science.aal2469} {\bibfield  {journal} {\bibinfo  {journal}
  {Science}\ }\textbf {\bibinfo {volume} {355}},\ \bibinfo {pages} {156}
  (\bibinfo {year} {2017})}\BibitemShut {NoStop}%
\bibitem [{\citenamefont {Mi}\ \emph {et~al.}(2018)\citenamefont {Mi},
  \citenamefont {Benito}, \citenamefont {Putz}, \citenamefont {Zajac},
  \citenamefont {Taylor}, \citenamefont {Burkard},\ and\ \citenamefont
  {Petta}}]{Mi2018}%
  \BibitemOpen
  \bibfield  {author} {\bibinfo {author} {\bibfnamefont {X.}~\bibnamefont
  {Mi}}, \bibinfo {author} {\bibfnamefont {M.}~\bibnamefont {Benito}}, \bibinfo
  {author} {\bibfnamefont {S.}~\bibnamefont {Putz}}, \bibinfo {author}
  {\bibfnamefont {D.~M.}\ \bibnamefont {Zajac}}, \bibinfo {author}
  {\bibfnamefont {J.~M.}\ \bibnamefont {Taylor}}, \bibinfo {author}
  {\bibfnamefont {G.}~\bibnamefont {Burkard}}, \ and\ \bibinfo {author}
  {\bibfnamefont {J.~R.}\ \bibnamefont {Petta}},\ }\href {\doibase
  10.1038/nature25769} {\bibfield  {journal} {\bibinfo  {journal} {Nature}\
  }\textbf {\bibinfo {volume} {555}},\ \bibinfo {pages} {599} (\bibinfo {year}
  {2018})}\BibitemShut {NoStop}%
\bibitem [{\citenamefont {Landig}\ \emph {et~al.}(2018)\citenamefont {Landig},
  \citenamefont {Koski}, \citenamefont {Scarlino}, \citenamefont {Mendes},
  \citenamefont {Blais}, \citenamefont {Reichl}, \citenamefont {Wegscheider},
  \citenamefont {Wallraff}, \citenamefont {Ensslin},\ and\ \citenamefont
  {Ihn}}]{Landig2018}%
  \BibitemOpen
  \bibfield  {author} {\bibinfo {author} {\bibfnamefont {A.~J.}\ \bibnamefont
  {Landig}}, \bibinfo {author} {\bibfnamefont {J.~V.}\ \bibnamefont {Koski}},
  \bibinfo {author} {\bibfnamefont {P.}~\bibnamefont {Scarlino}}, \bibinfo
  {author} {\bibfnamefont {U.~C.}\ \bibnamefont {Mendes}}, \bibinfo {author}
  {\bibfnamefont {A.}~\bibnamefont {Blais}}, \bibinfo {author} {\bibfnamefont
  {C.}~\bibnamefont {Reichl}}, \bibinfo {author} {\bibfnamefont
  {W.}~\bibnamefont {Wegscheider}}, \bibinfo {author} {\bibfnamefont
  {A.}~\bibnamefont {Wallraff}}, \bibinfo {author} {\bibfnamefont
  {K.}~\bibnamefont {Ensslin}}, \ and\ \bibinfo {author} {\bibfnamefont
  {T.}~\bibnamefont {Ihn}},\ }\href {\doibase 10.1038/s41586-018-0365-y}
  {\bibfield  {journal} {\bibinfo  {journal} {Nature}\ }\textbf {\bibinfo
  {volume} {560}},\ \bibinfo {pages} {179} (\bibinfo {year}
  {2018})}\BibitemShut {NoStop}%
\bibitem [{\citenamefont {Breeze}\ \emph {et~al.}(2017)\citenamefont {Breeze},
  \citenamefont {Salvadori}, \citenamefont {Sathian}, \citenamefont {Alford},\
  and\ \citenamefont {Kay}}]{Breeze2017}%
  \BibitemOpen
  \bibfield  {author} {\bibinfo {author} {\bibfnamefont {J.~D.}\ \bibnamefont
  {Breeze}}, \bibinfo {author} {\bibfnamefont {E.}~\bibnamefont {Salvadori}},
  \bibinfo {author} {\bibfnamefont {J.}~\bibnamefont {Sathian}}, \bibinfo
  {author} {\bibfnamefont {N.~M.}\ \bibnamefont {Alford}}, \ and\ \bibinfo
  {author} {\bibfnamefont {C.~W.~M.}\ \bibnamefont {Kay}},\ }\href {\doibase
  10.1038/s41534-017-0041-3} {\bibfield  {journal} {\bibinfo  {journal} {npj
  Quantum Information}\ }\textbf {\bibinfo {volume} {3}},\ \bibinfo {pages}
  {40} (\bibinfo {year} {2017})}\BibitemShut {NoStop}%
\bibitem [{\citenamefont {Fedorova}\ \emph {et~al.}(2020)\citenamefont
  {Fedorova}, \citenamefont {Qiu}, \citenamefont {Linden},\ and\ \citenamefont
  {Kroha}}]{Fedorova2020}%
  \BibitemOpen
  \bibfield  {author} {\bibinfo {author} {\bibfnamefont {Z.}~\bibnamefont
  {Fedorova}}, \bibinfo {author} {\bibfnamefont {H.}~\bibnamefont {Qiu}},
  \bibinfo {author} {\bibfnamefont {S.}~\bibnamefont {Linden}}, \ and\ \bibinfo
  {author} {\bibfnamefont {J.}~\bibnamefont {Kroha}},\ }\href {\doibase
  10.1038/s41467-020-17510-z} {\bibfield  {journal} {\bibinfo  {journal}
  {Nature Communications}\ }\textbf {\bibinfo {volume} {11}},\ \bibinfo {pages}
  {3758} (\bibinfo {year} {2020})}\BibitemShut {NoStop}%
\end{thebibliography}%
 %%%%%%%%%%%%%%%%%%%%%%%%%%%%%
 % Magnetic Systems + Curvature 
 %%%%%%%%%%%%%%%%%%%%%%%%%%%%%
 
\end{document}